\documentclass[12pt]{iopart}
\pdfoutput=1 
\usepackage{setspace}
\usepackage{tocloft}
\expandafter\let\csname equation*\endcsname\relax
\expandafter\let\csname endequation*\endcsname\relax
\usepackage{amsmath}
\usepackage{amsfonts,amsthm,color}	
\usepackage[pdftex]{graphicx}						
\usepackage{url}
\usepackage{enumitem}

\def\dd{{{\mathrm d}}}

\def\betamin{{\beta_{\rm min}}}
\def\betamax{{\beta_{\rm max}}}
\newcommand{\ud}{\,\mathrm{d}}

\usepackage{xcolor}
\providecommand{\keywords}[1]{\textbf{Keywords} #1}

\begin{document} 

\title[Simulated Tempering Method in the Infinite Switch
Limit]{Simulated Tempering Method in the Infinite Switch Limit with
  Adaptive Weight Learning}

\author{Anton Martinsson$^1$, Jianfeng Lu$^2$ , Benedict
  Leimkuhler$^1$, Eric Vanden-Eijnden$^3$} 

\address{$^1$ School of Mathematics and
  Maxwell Institute, University of Edinburgh, James Clerk Maxwell
  Building, Edinburgh EH9 3FD, United Kingdom} 
\address{$^2$ Departments
  of Mathematics, Physics, and Chemistry, Duke University, Durham NC
  27708, USA} 
\address{$^3$ Courant Institute for the Mathematical
  Sciences, New York University, 251 Mercer street, New York, NY
  10012, USA } 
\ead{Anton.Martinsson@ed.ac.uk, jianfeng@math.duke.edu,
  B.Leimkuhler@ed.ac.uk, and eve2@cims.nyu.edu}

\begin{abstract}
  We investigate the theoretical foundations of the simulated
  tempering (ST) method and use our findings to design an efficient
  accelerated sampling algorithm.  Employing a large deviation
  argument first used for replica exchange molecular dynamics [Nuria
  \textit{et al.} {\em J. Chem. Phys.} \textbf{135}:134111 (2011)], we
  demonstrate that the most efficient approach to simulated tempering
  is to vary the temperature infinitely rapidly.  In this limit, we
  can replace the equations of motion for the temperature and physical
  variables by averaged equations for the latter alone, with the
  forces rescaled according to a position-dependent function defined
  in terms of temperature weights. The averaged equations are similar
  to those used in Gao's integrated-over-temperature method, except
  that we show that it is better to use a continuous rather than a
  discrete set of temperatures.  We give a theoretical argument for
  the choice of the temperature weights as the reciprocal partition
  function, thereby relating simulated tempering to Wang-Landau
  sampling. Finally, we describe a self-consistent algorithm for
  simultaneously sampling the canonical ensemble and learning the
  weights during simulation. This infinite switch simulated tempering
  (ISST) algorithm is tested on three examples of increasing
  complexity: a system of harmonic oscillators; a continuous variant
  of the Curie-Weiss model, where ISST is shown to perform better than
  standard ST and to accurately capture the second-order phase
  transition observed in this model; and Alanine-12 in vacuum, where
  ISST also compares favorably with standard ST in its ability to
  calculate the free energy profiles of the root mean square deviation
  (RMSD) and radius of gyration of the molecule in the 300-500K
  tempearture range.
\end{abstract}


\maketitle



 
\section{Introduction}
\label{sec:intro}

The sampling of probability distributions in high dimensions is a
fundamental challenge in computational science, with broad
applications to physics, chemistry, biology, finance, machine
learning, and other areas
(e.g.~\cite{frenkel2001understanding,MacKay2003InformationTI,
  richard2007efficient,landau2014guide}). The methods of choice to
tackle this problem are based on Monte Carlo or the simulation of
stochastic differential equations such as the Langevin equation,
either of which can be designed to be ergodic for a wide class of
probability distributions. Yet, straightforward application of these
methods typically fails when the distribution displays complex
features such as multimodality.  In high dimensions, the per-step cost
of generating samples is significant and can be taken as the unit of
computational effort; naive approaches may require many millions or
billions of iterations.  A typical case in point arises in
computational chemistry, where molecular dynamics (MD) simulation has
become an invaluable tool for resolving chemical structures, exploring
the conformational states of biomolecules and computing free
energies. Yet, despite its versatility, the use of MD simulation is
often limited by the intrinsic high dimensionality of the systems
involved and the presence of entropic and energetic barriers which
lead to slow diffusion and necessitate the use of very long
trajectories. A typical MD simulation is thus spent oversampling a few
free energy minima, with consequent poor approximation of properties
of interest.

Numerous methods have been introduced for overcoming the intrinsic
complexities of high-dimensional sampling. These accelerated sampling
method include the Wang-Landau method \cite{WaLa2001,JuPeVo2014} which
directly estimates the density of states and thus the entropy during
simulation; metadynamics \cite{LaPa2002,BaBuPa2008} which
progressively modifies the potential energy during simulation to
flatten out the landscape and accelerate transitions between states;
temperature-accelerated molecular dynamics (TAMD)
\cite{MaVa2006,AbVa2010}, which effectively evolves collective
variables on their free energy landscape at the physical temperature
but use an artificially high temperature to speed-up their dynamics; or
the adaptive biasing force (ABF) method \cite{DaRoPo2008} which
modifies forces using a continually refined estimated free energy
gradient.

Another popular class of accelerated sampling schemes is based on
adding the temperature to the system state variables and allowing it
to vary during the simulation. These methods, which originated from
simulated annealing, were first introduced for Langevin
dynamics\cite{Aluffi-Pentini1985}.  They include replica exchange
molecular dynamics (REMD)~\cite{Swendsen1986,Geyer1991,%
  Neal1996,Hansmann1997,Sugita1999}, in which several replicas of the
system are evolved at different temperatures which they exchange, and
simulated tempering (ST) methods, in which the temperature is treated
as a dynamical variable evolving in tandem with the physical variables
\cite{Marinari1992,Hansmann1997a}. The general idea underpinning REMD
and ST is that modification of the temperature introduces nonphysical
states that accelerate the dynamics of the physical system at target
conditions. Closely related to the tempering methods are various
``alchemical'' simulation schemes which allow dynamical modification
of parameters of the energy function describing the molecule, which are
in spirit similar to what is done in umbrella sampling.  Hamiltonian replica exchange is
an example, for example \cite{Banba2000,Woods2003,%
  Bitetti-Putzer2003,Okamoto2004,FARALDO-GOMEZ2007,Hritz2008} which
involves e.g. softening a dihedral bend \cite{Okazaki2006,Miao2014} or
reducing the forces acting between protein and explicit solvent bath
\cite{Liu2005}.

In this paper, we focus our attention primarily on the original ST in
which the temperature is evolved along with the particle positions
and momenta. In the standard implementations of the method, the
temperature is treated as a discrete random variable, evolving
together with the physical variables via a discrete-time Markov
chain. The practitioner is normally required to make a choice of say
$M$ temperatures distributed over some range,
\begin{equation}
  T_{\rm min}=T_1<T_2<\ldots<T_M=T_{\rm max}, \label{eq:1}
\end{equation}
which defines the temperature ``ladder''.  Letting
$\beta_i = (k_B T_i)^{-1}$, we prescribe a weight $\omega(\beta_i)>0$ at each of
these reciprocal temperatures, and the system
state variables are then evolved along with the reciprocal temperature
in the following way:\footnote{%
  Note that in this version, tempering amounts to rescaling the forces
  rather than actually changing the temperature of the bath. In the
  context of MD simulations, this is the way ST is typically
  implemented in practice, to not temper with the kinetic energy of
  the system, see~\eqref{eq:rho_xpbeta}.}
\begin{enumerate}
\item Given the current state of the reciprocal temperature, say
  $\beta_i$, standard MD simulations are performed with the force
  rescaled by the factor $\beta_i/\beta$ for a lag time of duration
  $\tau>0$ (here $\beta$ is fixed);
\item At the end of each time interval of duration $\tau$, a switch
  from $\beta_i$ to some $\beta_j\not=\beta_i$ is attempted, and
  accepted or rejected according to a Metropolis-Hastings criterion with
  acceptance probability
\begin{equation}
  \alpha_{i j} = \frac{\omega(\beta_i)}{\omega(\beta_j)} e^{-(\beta_i - \beta_j) V(q) }.
	\label{eq:ST}
\end{equation}
Here $V(q)$ is the potential energy at the current position
$q \in \mathcal{D}$.
\end{enumerate}
\noindent
The simple idea in ST is that exploration is aided by the high
temperatures, when $\beta_i<\beta$, since the rescaling of the force
by the factor $\beta_i/\beta<1$ will help the system traverse
energetic barriers easily. The lower temperatures, when
$\beta_i>\beta$, complement the sampling by providing enhanced
resolution of low energy states. The scheme above guarantees that this
acceleration of the sampling is done in a controlled way, in that we
know that the ST dynamics is ergodic with respect to the extended
Gibbs distribution
\begin{equation}
  \rho(q,p,\beta_i) = C^{-1}(\beta) \omega(\beta_i)
  e^{-\frac12\beta p^Tm^{-1} p- \beta_i V(q)},
  \qquad q\in \mathcal{D}, \ \ p \in \mathbb{R}^d, \ \ i=1\ldots, M
  ~,
  \label{eq:rho_xpbeta0}
\end{equation}
where $m$ is the mass tensor, $d$ is the physical dimension times the number
of particles, and $C(\beta)$ is the normalization constant
\begin{equation}
  \label{eq:2}
  C(\beta) = \sum_{i=1}^M \int_{\mathcal{D} \times \mathbb{R}^d}
  \omega(\beta_i) e^{-\frac12\beta p^Tm^{-1} p- \beta_i V(q)}\dd q \dd p
  ~.
\end{equation}
Knowledge of~(\ref{eq:rho_xpbeta0}) permits one to unbias the ST sampling and compute averages with respect to the target Gibbs
distribution in the original system state space:
\begin{equation}
  \rho_\beta (q,p) = Z^{-1}(\beta) e^{-\frac12\beta  p^Tm^{-1} p- \beta V(q)},
\label{eq:gibbs}
\end{equation}
where
$Z(\beta) = \int_{\mathcal{D} \times \mathbb{R}^d} \exp\left[
  -\tfrac12 \beta p^Tm^{-1} p- \beta V(q) \right] \dd q \dd p $ is the
partition function. Standard techniques, such as the Langevin Dynamics
associated with \eqref{eq:gibbs}, explore this measure poorly. ST can
in principle accelerate sampling for the reasons listed
above. However, to design an effective implementation of ST,
practitioners must make choices for the temperature ladder
in~(\ref{eq:1}), switching frequency $\tau$, and weight factor
$\omega(\beta)$ in \eqref{eq:ST}. These choices are all non-trivial
and exhibit some clear interdependence. Our aim here is to explain how
to chose these parameters and show how to do so in
practice. Specifically:

\begin{itemize}[leftmargin=.2in]
\item By adapting the large deviation approach proposed by Dupuis
  \textit{et al.} in Refs.~\cite{Plattner2011,Dupuis2012} in the
  context of REMD, we show that ST is most efficient if operated in
  the infinite switch limit, $\tau\to0$. In this limit, one can derive
  a limiting equation for the particle positions and momenta alone, in
  which the original potential is replaced by an averaged
  potential. In the context of the standard ST method described above,
  this averaged potential reads
  \begin{equation}
    \label{eq:3discrete}
    \bar V(q) = -\beta^{-1} \log \sum_{i=1}^M \omega(\beta_i)
    e^{-\beta_i V(q)}.
  \end{equation}
  In the infinite switch limit, the ST method then becomes similar to the
  ``integrate-over-tem\-pe\-ra\-ture'' method that Gao proposed in
  Ref.~\cite{Gao2008}, and there is no longer any need to update
  the temperatures -- they have been averaged over.
\item Regarding the choice of temperature ladder, we show that it
  is better to make the reciprocal temperature vary continuously
  between $\beta_{{\rm min}}$ and $\beta_{{\rm max}}$. In this case,
  the averaged potential~\eqref{eq:3discrete} becomes
\begin{equation}
    \label{eq:3}
    \bar V(q) = -\beta^{-1} \log
    \int_{\beta_{{\rm min}}}^{\beta_{{\rm max}}}
    \omega(\beta_c)
    e^{-\beta_c V(q)} d\beta_c
    ~,
  \end{equation}
  and we can think of~(\ref{eq:3discrete}) as a way to approximate
  this integral by discretization.  When the reciprocal temperature
  takes continuous values, the infinite switch limit of ST is also the
  scheme one obtains by infinite acceleration of the temperature
  dynamics in the continuous tempering method proposed in
  Ref.~\cite{Gobbo2015}.
\item Regarding the choice of $\omega(\beta_i)$, the conventional wisdom
  is to take $\omega(\beta_i) = Z_q^{-1}(\beta_i)$, where
  $Z_q^{-1}(\beta)$ is the configurational part of the partition
  function:
\begin{equation}
  \label{eq:8}
  Z_q(\beta) = \int_{\mathcal{D}}  e^{-\beta  V(q)} dq~.
\end{equation}
This choice is typically justified because it flattens the marginal of
the distribution~(\ref{eq:rho_xpbeta0}) over the temperatures. Indeed
this marginal distribution is given for $i=1,\ldots,M$ by
\begin{equation}
  \label{eq:11}
  p(\beta_i) = C^{-1}(\beta) \int_{\mathcal{D} \times \mathbb{R}^d}
  \omega(\beta_i) e^{-\frac12\beta p^Tm^{-1} p- \beta_i V(q)}\dd q \dd
  p =\frac{\omega(\beta_i)
  Z_q(\beta_i)}{\sum_{j=1}^M\omega(\beta_j) Z_q(\beta_j)}
	~,
\end{equation}
which is uniform if $\omega(\beta_i) = Z_q^{-1}(\beta_i)$. Here we
show that the choice $\omega(\beta_i) = Z_q^{-1}(\beta_i)$ also
flattens the distribution of potential energy in the modified ensemble
with averaged potential~(\ref{eq:3}). Interestingly, this offers a new
explanation of why ST becomes inefficient if the system undergoes a phase
transition, such that its density of states is not log-concave. This
perspective will allow us to make a connection between ST and the
Wang-Landau method \cite{WaLa2001,JuPeVo2014}.

\item Building on these results, an additional contribution of this
  article is to give a precise formulation of an algorithm for
  learning the weights on the fly. The implicit coupling between
  physical dynamics and weight determination complicates
  implementation, and remains an active area of
  research~\cite{Park:2007id,Gao2008,Roberts2014min,tan2017,Yang:2018wy}. This
  problem of weight determination is comparable to a machine learning
  problem in which parameters of a statistical model must be inferred
  from data (in this case the microscopic trajectories
  themselves). This algorithm derives from an estimator of the
  partition function that utilizes a full MD trajectory over the full
  set of temperatures $\beta_c\in[\betamin,\betamax]$. This is similar
  to the Rao-Blackwellization procedure proposed in
  Ref.~\cite{Carlson:2016} (see also~\cite{Madras1999importance}) and
  in contrast to, but more efficient than, traditional methods which
  are restricted to a particular temperature $\beta_c$.
\end{itemize}

As was outlined above, to establish these results it will be convenient to
work with a continuous formulation of ST, in which the reciprocal
temperatures vary continuously both in time and in value in the range
$[\beta_{{\rm min}},\beta_{{\rm max}}]$. This formulation is
introduced next, in Sec.~\ref{sec:newST}. We stress that it
facilitates the analysis, but does not affect the results: \textit{all
  our conclusions also hold if we were to start from the standard ST
  algorithm described above.} 

The remainder of this paper is organized as follows: In
Sec.~\ref{sec:isst} we present the theoretical foundations of ST using
the continuous variant that we propose, which is introduced in
Sec.~\ref{sec:newST}. In Sec.~\ref{sec:LDP} we derive a closed
effective equation for the system state variables alone in the
infinite switch limit and we justify that it is most efficient to
operate ST in this limit. In Sec.~\ref{sec:expect} we show how to
estimate canonical expectations using this limiting equation -- the
same estimator can also be used for standard ST and amount to
performing a Rao-Blackwellization of the standard estimator used in
that context. We also show how to estimate the partition function
$Z_q(\beta)$. In Sec.~\ref{sec:DOS} we go on to explain why the choice
$\omega(\beta) = Z_q^{-1}(\beta)$ is optimal.  Finally, in
Sec.~\ref{sec:learning} we explain how to learn these weights on the
fly. These theoretical results are then used to develop a practical
numerical scheme in Sec.~\ref{sec:Disc}, and this scheme is tested on
thhree examples in Sec.~\ref{sec:NumExp}: the $d$-dimensional harmonic
oscillator, which allows us to investigate the effects of
dimensionality in a simple situation where all the relevant quantities
can be calculated analytically; a continuous version of the
Curie-Weiss model, which displays a second-order phase transition and
allows us to investigate the performance of ST in the infinite
temperature switch limit when this happens; and finally the Alanine-12
molecule in vacuum, where we use ISST to calculate the free energy of
the root mean square deviation (RMSD) and radius of gyration Rg of the
molecule to investiagte its conforamtioanl states in the 300-500K
temperature range. In these last two examples, we also compare the
performances of ISST to those of standard ST, and observe that the
former performs significantly betrer than the latter. Finally, some concluding
remarks are made in Sec.~\ref{sec:conclu}. 

\section{Foundations of Infinite Switch Simulated Tempering (ISST)}
\label{sec:isst}
In this section, we discuss the theoretical foundation of simulated
tempering, in particular deriving a simplified system of equations for
the physical variables that eliminates the need to perform a discrete
switching over temperature.  Although we still technically work with a
temperature `ladder', as in other works in this area, we shall see that
these are only used to perform the averaging across temperatures in an
efficient way.

\subsection{A Continuous formulation of Simulated Tempering}
\label{sec:newST}

In order to simplify our presentation and advance the large
deviation argument, we first replace standard simulated tempering,
where $\beta_i$ is taken from a discrete sequence in $\betamin$ to
$\betamax$ with a model that incorporates a continuously variable
reciprocal temperature $\beta_c$, taking values in the interval
$[\betamin,\betamax]$. Note that $\beta_c$, which continuously varies, should not
be confused with the physical reciprocal temperature $\beta$, which is
fixed. In this continuous tempering setting, the extended Gibbs
distribution has density
\begin{equation}
  \rho(q,p,\beta_c) = C^{-1}(\beta) \omega(\beta_c) e^{-\tfrac12 \beta p^T
    m^{-1} p -\beta_c  V(q)}, \qquad q\in \mathcal{D}, \ \ p \in
  \mathbb{R}^d, \ \ 
  \beta_c\in[\beta_{{\rm min}},\beta_{{\rm max}}]
  ~,
  \label{eq:rho_xpbeta}
\end{equation}
where $C(\beta)$ is a normalization constant:
\begin{equation}
  \label{eq:Cnormdef}
  \begin{aligned}
    C(\beta) & = \int_{\betamin}^{\betamax} \int_{\mathcal{D} \times
      \mathbb{R}^d} \omega(\beta_c) e^{-\tfrac12 \beta p^T m^{-1} p
      -\beta_c V(q)} \dd q \dd p \dd \beta_c ~,\\
    & = (2\pi \beta^{-1})^{d/2} (\det m)^{1/2}\int_{\betamin}^{\betamax}
    \omega(\beta_c) Z_q(\beta_c) \dd \beta_c~.
  \end{aligned}
\end{equation}

For sampling purposes, we also need to introduce a dynamical system
that is ergodic with respect to the
distribution~\eqref{eq:rho_xpbeta}. A possible choice is
\begin{equation}
  \begin{split}
    \dot q =& m^{-1} p ~,
    \\ \dot p =& - \beta^{-1} \beta_c \nabla V - \gamma p + 
    \sqrt{2 \gamma \beta^{-1} m^{-1} } \, \eta_p ~,
    \\ \alpha \dot \beta_c =& \beta^{-1} V(q) - \beta^{-1}
    \omega^{-1}(\beta_c) \omega'(\beta_c)
    + \sqrt{2\beta^{-1} \alpha }\,\eta_{\beta_c} ~.
		\label{eq:continuous}
	\end{split}
\end{equation}
Here $\gamma$ is a Langevin friction coefficient, $\eta_p$ and
$\eta_{\beta_c}$ represent independent white noise processes, $\alpha$
is a time-scale parameter, and we recall that $\beta_c$ is the
tempering variable that evolves whereas $\beta$ is the physical
reciprocal temperature that is fixed. Note that other choices of
dynamics are possible \cite{Gobbo2015}, as long as they are ergodic
with respect to~(\ref{eq:rho_xpbeta}); working in the limit where $\beta_c$ is
infinitely fast compared
to $(q,p)$ can be applied to these other dynamical systems as
well. The effective equation one obtains in that limit
is~\eqref{eq:continuouslim}, given that the system variables $(q,p)$
are the same as in~\eqref{eq:continuous} (i.e. the specifics of how
$\beta_c$ evolves does not matter in this limit).

\subsection{Infinite Switch Limit}
\label{sec:LDP}

In this subsection, we argue that it is best to use simulated tempering in an infinite
switch limit, and we derive an effective equation for the physical
state variables that emerge in that limit. In the context
of~(\ref{eq:continuous}), this infinite switch limit can be achieved by
letting $\alpha\to0$ in this equation, in which case $\beta_c$
equilibrates rapidly on the $O(\alpha)$ timescale before $(q,p)$
moves. The state variables $(q,p)$ thus only feel the average effect
of $\beta_c$ on the $O(1)$ timescale, that is, (\ref{eq:continuous})
reduces to the following limiting system for $(q,p)$ alone
\begin{equation}
  \begin{split}
    \dot q =& m^{-1} p ~, \\ \dot p =& - \beta^{-1} \bar \beta(V(q))
    \nabla V - \gamma p + \sqrt{2 \gamma \beta^{-1} m^{-1} } \,
    \eta_p ~,	\label{eq:continuouslim}
	\end{split}
\end{equation}
where $\bar \beta(V(q)) $ is the conditional average of $\beta_c$ with
respect to~(\ref{eq:rho_xpbeta}) at $(q,p)$ fixed:
\begin{equation}
  \label{eq:4}
  \bar \beta(V(q)) = \frac{\int_{\betamin}^{\betamax} \beta_c
    \omega(\beta_c) e^{-\beta_c  V(q)} \dd
    \beta_c }{\int_{\betamin}^{\betamax} \omega(\beta_c) e^{-\beta_c  V(q)} \dd
    \beta_c}
	~,
\end{equation}
(\ref{eq:continuouslim}) can be derived by standard averaging theorems
for Markov processes, and it is easy to see that in this equation we
can view the effective force as the gradient of a modified potential:
\begin{equation}
  \label{eq:5a}
  \beta^{-1} \bar{\beta}(V(q))
    \nabla V = \nabla \bar{V}(q)
    ~,
\end{equation}
where $\bar{V}(q)$ is the averaged potential defined
in~(\ref{eq:3}). We will analyze the properties of this effective
potential in more details later in Sec.~\ref{sec:DOS}. We stress
that~\eqref{eq:4} is a closed equation for $(q,p)$. In other words, in
the infinite switch limit it is no longer necessary to evolve the
reciprocal temperature $\beta_c$: the averaged effect the dynamics of
$\beta_c$ has on $(q,p)$ is fully accounted for
in~\eqref{eq:continuouslim}.

Let us now establish that the limiting equations in~\eqref{eq:continuouslim} are
more efficient than the original (\ref{eq:continuous}) (or variants
thereof that lead to the same limiting equation) for sampling
purposes. To this end we use the approach based on large deviation
theory proposed by Dupuis and collaborators
\cite{Plattner2011,Dupuis2012}. Define the empirical measure $\nu_T$
for the dynamics up to time $T$ by
\begin{equation}
  \label{eq:empiricalm}
\nu_T(q, p, \beta_c) = \frac{1}{T} \int_0^T 
\delta(q - q(t)) \delta( p - p(t)) \delta(\beta_c - \beta_c(t)) \ud t.
\end{equation} 
Donsker-Varadhan theory \cite{DonskerVaradhan:1975,
	DeuschelStroock:1989} states that as $T \to \infty$, the empirical
measure satisfies a large deviation principle with rate functional
given by
\begin{equation}\label{eq:DVaction}
  I_{\alpha}(\mu) = \sup_{g \in C_b^{\infty}(\mathbb{R}^{2N}; [1,
    \infty))} \int 
  - \frac{\mathcal{L}_{\alpha} g}{g} \ud\mu 
\end{equation}
where we denote by $\mathcal{L}_{\alpha}$ the infinitesimal generator
of \eqref{eq:continuous}:
\begin{equation}
\begin{aligned}
  \mathcal{L}_{\alpha} f & = \frac{p}{m} \partial_q f -
  \frac{\beta_c}{\beta} \partial_q V \cdot \partial_p f 
  - \gamma p \cdot \partial_p f + \gamma \beta^{-1}  m \Delta_p f \\
  & \qquad + \frac{1}{\alpha} \Bigl( \beta^{-1} \bigl(V
  - \partial_{\beta_c}\ln \omega(\beta_c)\bigr) 
  \cdot \partial_{\beta_c} f  
  + \beta^{-1} \Delta_{\beta_c} f \Bigr). 
\end{aligned}
\end{equation}
Colloquially, the large deviation principle asserts that the
probability that the empirical measure $\nu_T$ be close to $\mu$, is
asymptotically given for large $T$ by
\begin{equation}
  \label{eq:34}
  \mathbb{P}(\nu_T \approx \mu) \asymp \exp\left(- T^{-1} I_\alpha (\mu)\right)
  ~.
\end{equation}
Since, as we will see, $I_\alpha (\mu)=0$ if and only if $\mu$ is the
invariant measure associated with~\eqref{eq:continuous}, \eqref{eq:34}
gives an estimate of the (un)likelihood that $\nu_T$ is different from
this invariant measure when $T$ is large.


For Langevin dynamics, the rate functional can be further simplified
\cite{Wu:2001, BodineauLefevere:2008}, as we show next. Due to the
Hamiltonian part, the generator $\mathcal{L}_{\alpha}$ is not
self-adjoint with respect to $L^2(\rho)$, where $\rho$ is the density
in~(\ref{eq:rho_xpbeta0}); denote by $\mathcal{L}_{\alpha}^{\ast}$ the
weighted adjoint, then we have
\begin{equation}
\mathcal{L}_{\alpha}^{\ast} = \Pi \mathcal{L}_{\alpha} \Pi, 
\end{equation}
where $\Pi$ is the operator that flips the momentum direction:
$\Pi(q, p, \beta_c) = (q, -p, \beta_c)$. We will split the generator
into the symmetric part (corresponding to damping and diffusion in
momentum), the anti-symmetric part (corresponding to Hamiltonian
dynamics), and the tempering part (corresponding to the dynamics of
$\beta_c$):
\begin{equation}
\mathcal{L}_{\alpha} = \mathcal{L}^S + \mathcal{L}^H + \alpha^{-1} \mathcal{L}^{T}
~,
\end{equation}
with 
\begin{align}
  \mathcal{L}^S f & = - \gamma p \cdot \partial_p f + \gamma \beta^{-1} m \Delta_p f~, \\
  \mathcal{L}^H f & =  \frac{p}{m} \partial_q f - 
                    \frac{\beta_c}{\beta} \partial_q V \cdot \partial_p f~, \\
  \mathcal{L}^T f & =  \beta^{-1} \bigl(V - \partial_{\beta_c}\ln \omega(\beta_c)\bigr) \cdot \partial_{\beta_c} f  
                    + \beta^{-1} \Delta_{\beta_c} f~.
\end{align}
Let $f = \ud \mu / \ud \varrho$ be the Radon-Nikodym derivative of
$\ud \mu$ with respect to $\ud \varrho$ , where $\ud \varrho$ is the measure
with density~(\ref{eq:rho_xpbeta}) (i.e. the invariant measure for the
dynamics in~(\ref{eq:continuous})). If
$\int \lVert \partial_p f^{1/2} \rVert_2 \ud \varrho < \infty$, then
the large deviation rate functional is given by
\begin{equation}\label{eq:Imbeta}
\begin{aligned}
  I_{\alpha}(\mu) & =  - \int f^{1/2} \mathcal{L}^S f^{1/2} \ud \varrho \\
  & \hspace{0.5in} - \inf_g \; \Biggl( \frac{1}{4 \beta_{\ast}} \int
  \gamma m \lvert \partial_p g \rvert^2 \ud \mu + \frac{1}{2} \int
  \mathcal{L}^H g \ud \mu \Biggr) \\
  & \hspace{0.5in} - \frac{1}{2\alpha} \int f^{1/2} \mathcal{L}^T f^{1/2} \ud \varrho.
\end{aligned}
\end{equation}
In particular, after an integration by parts, the only $\alpha$-dependent term is given by
\begin{equation}
  - \frac{1}{2\alpha} \int f^{1/2} \mathcal{L}^T f^{1/2} \ud \rho
  = \frac{1}{2\alpha} \int \bigl\lvert \nabla_{\beta_c} f^{1/2} \bigr\rvert^2 \ud \varrho.
\end{equation}
It is thus clear that for $\alpha < \alpha'$, we have
$I_{\alpha}(\mu) \geq I_{\alpha'}(\mu)$, which leads to the
conclusion that the rate function $I_{\alpha}$ is pointwise
monotonically decreasing  in~$\alpha$. 

In summary, to increase the large deviation rate functional for the
empirical measure~\eqref{eq:empiricalm} (and hence to have faster
convergence), we should take a smaller $\alpha$. This ultimately justifies taking
the infinite switch limit $\alpha \to 0$ of the evolution equations 
in~(\ref{eq:continuous}), in which case they reduce
to~(\ref{eq:continuouslim}).

\subsection{Estimation of Canonical Expectations and $Z_q(\beta)$}
\label{sec:expect}

It is easy to see that~(\ref{eq:continuouslim})
(similar to~\eqref{eq:continuous} if we only process $(q(t),p(t))$) is
ergodic with respect to the density obtained by
marginalizing~(\ref{eq:rho_xpbeta}) on $(q,p)$:
\begin{equation}
  \label{eq:7}
  \bar \rho (q,p) =  C^{-1}(\beta) e^{-\frac12 \beta p^T m^{-1} p-\beta
    \bar V(q)} =  C^{-1}(\beta) e^{-\frac12 \beta p^T m^{-1} p} 
  \int_{\betamin}^{\betamax} \omega(\beta_c) e^{-\beta_c  V(q)} \dd
  \beta_c,
\end{equation}
where $C(\beta)$ is the normalization constant defined
in~(\ref{eq:Cnormdef}). As a result, if $A(q)$ is an observable of
interest, its canonical expectation at temperature $\beta_c$ can be
expressed as a weighted average,
\begin{equation}
\begin{split}
	\mathbb{E}_{\beta_c} \left[ A \right] 
	=& \int_{\mathcal{D} \times \mathbb{R}^d } A(q) \rho_{\beta_c}(q,p) \dd q \dd p
	\\=& \int_{\mathcal{D} \times \mathbb{R}^d } A(q) W_{\beta_c}
        (q) \bar \rho(q,p) \dd q \dd p
	\\=& \lim_{T \to \infty} \frac{1}{T} \int_{0}^{T} A(q(t)) W_{\beta_c} (q(t)) \dd t ,
	\label{eq:reweight}
\end{split}
\end{equation}
where we assumed ergodicity in the last equality and we define,
\begin{equation}
  \label{eq:6}
  \begin{aligned}
    W_{\beta_c} (q) &= \frac{\rho_{\beta_c}(q, p)} {\bar{\rho}(q, p)} \\
    & = Z^{-1}_q(\beta_c) \frac{\int_{\betamin}^{\betamax}
      \omega(\beta'_c) Z_q(\beta'_c) d\beta'_c}
    {\int_{\betamin}^{\betamax} \omega(\beta'_c) e^{-(\beta'_c -\beta_c)
        V(q)} d\beta'_c} 
    ~.
  \end{aligned}
\end{equation}
Expression~\eqref{eq:reweight} is different from the standard approach used
in ST in that it uses the data from all
$\beta_c \in [\betamin,\betamax]$ to calculate expectations at
reciprocal temperature $\beta$.  By contrast, the typical estimator
used in ST only uses the part of the time series $(q(t),p(t))$ during
which $\beta_c(t) = \beta $ (or $\beta_i=\beta$ when one uses a
discrete set of reciprocal temperatures). As identified in
Ref.~\cite{Carlson:2016}, \eqref{eq:reweight} amounts to performing
a Rao-Blackwellization of the standard ST estimator, which always
reduces the variance of this estimator.

In the same way, the expression~\eqref{eq:6} for the weights in~\eqref{eq:reweight} clearly indicates that the reweighting can
only be done with knowledge of $Z_q(\beta)$, which is typically not available \textit{a~priori}. Often the aim of sampling is precisely to
calculate $Z_q(\beta)$. Thus to make~(\ref{eq:continuouslim}) useful
one also needs to design a way to estimate $Z_q(\beta)$ from the
simulation; this issue also arises with standard ST or when one
uses~(\ref{eq:continuous}). This can be done using the
following estimator that can be verified by direct calculation using
the definition of $\bar\rho(q,p)$ in~\eqref{eq:7},
\begin{equation}
  \label{eq:Zcestimator}
  \begin{split}
    \frac{Z_q(\beta)}{\int_{\betamin}^{\betamax} Z_q(\beta_c)
      \omega(\beta_c) \dd \beta_c} =& \int_{\mathcal{D} \times
      \mathbb{R}^{d} } \frac{ e^{-\beta V(q)} \bar\rho(q,p) \dd
      q \dd p}{\int_{\betamin}^{\betamax} \omega(\beta_c)
      e^{-\beta_c V(q)} \dd \beta_c } \\=& \lim_{T \to \infty}
    \frac{1}{T} \int_{0}^{T} \frac{ e^{-\beta V(q(t))}
    }{\int_{\betamin}^{\betamax} \omega(\beta_c) e^{-\beta_c
        V(q(t))} \dd \beta_c } \dd t~,
  \end{split}
\end{equation}
where the last equality follows from ergodicity.  The
estimator~\eqref{eq:Zcestimator} permits the calculation of the ratio
$Z_q(\beta)/Z_q(\beta')$ for any pair $\beta$, $\beta'$. It will allow
us to kill two birds with one stone, since the scheme is most
efficient if used with $\omega(\beta)$ proportional to
$Z_q^{-1}(\beta)$.  By learning $Z_q(\beta)$ we are also
able to adjust $\omega(\beta)$ on-the-fly and thereby improve sampling
efficiency along the way. As mentioned before,
$\omega(\beta) \propto Z_q^{-1}(\beta)$ leads to a flattening of the
distribution of potential energy, as in
Wang-Landau sampling \cite{WaLa2001}, as discussed in the next subsection.

\subsection{Optimal Choice of the Temperature Weights}
\label{sec:DOS}

The choice $\omega(\beta) \propto Z_q^{-1}(\beta)$ is typically
justified because it flattens the marginal distribution
of~\eqref{eq:rho_xpbeta} with respect to reciprocal
temperature. Indeed, this marginal density is given by (which is the
continuous equivalent of~\eqref{eq:11}):
\begin{equation}
  \label{eq:14}
  p(\beta_c) = \int_{\mathcal{D}\times \mathbb{RR}^d} \bar
  \rho(q,p) \ud q
  \ud p = \frac{\omega(\beta_c) Z_q(\beta_c)
  }{\int_{\beta_{{\rm min}}}^{\beta_{{\rm max}}} \omega(\beta_c')
    Z_q(\beta_c') \ud\beta_c'}, \qquad \beta_c\in[\beta_{{\rm min}},\beta_{{\rm max}}].
\end{equation}
Having a flat $p(\beta_c)$ is deemed advantageous since it guarantees
that the reciprocal temperature explores the entire available range
homogeneously and does not get trapped for long stretches of time in
regions of $[\beta_{{\rm min}},\beta_{{\rm max}}]$ where $p(\beta_c)$
is low. In the infinite switch limit, however, the reciprocal
temperature is never trapped on the $O(1)$ time-scale over which
$(q,p)$ varies, since the reciprocal temperature evolves infinitely
fast and is averaged over. It is therefore useful to give an
alternative justification for the choice
$\omega(\beta) \propto Z^{-1}_c(\beta)$.

Such a justification can be found by looking at the probability
density function of the potential energy $V(q)$ in the system governed
by (\ref{eq:continuous}) or its limiting
version~(\ref{eq:continuouslim}). It is given by,
\begin{equation}
  \label{eq:rhocE}
  \bar \rho(E)
  = \int_{\mathcal{D}\times\mathbb{R}^d}  \delta( V(q) - E)
  \bar\rho(q,p) \dd q \dd p
  =\frac{\int_{\betamin}^{\betamax}e^{-\beta_c E} \omega(\beta_c)
    \dd\beta_c}
  { \int_{\betamin}^{\betamax} Z_q(\beta_c) \omega(\beta_c) \dd\beta_c}
  \, \Omega(E),
\end{equation}
where $\Omega (E)$ is the density of states
\begin{equation}
  \label{eq:5}
  \Omega(E) = \int_{\mathcal{D}} \delta( V(q) - E ) \dd q.
\end{equation}
Next we show that, in the limit of large system size, $\bar \rho(E)$
can be made flat for a band of energies by setting
$\omega(\beta) \propto Z_q^{-1}(\beta)$. Note that, in contrast, the
probability density of $V(q)$ of the original system with canonical
density $\rho_{\beta}(q, p)$, i.e.,
\begin{equation}
  \rho_{\beta}(E)
  = \int_\mathcal{D} \delta( V(q) - E) \rho_{\beta}(q,p) \dd q \dd p 
  =Z_q^{-1}(\beta)e^{-\beta E }\,  \Omega(E)~,
\end{equation}
is in general very peaked at one value of $E$ in the large system size
limit; this will also become apparent from the argument below.

To begin, use the standard expression of $Z_q(\beta)$ in terms of $\Omega(E)$
\begin{align}
  Z_q(\beta)
  &= \int_{\mathcal{D}} e^{-\beta V(q) }\dd q \nonumber\\
  &= \int _{\mathcal{D}} \int_{0}^\infty \delta(E - V(q)) \dd E
    e^{-\beta V(q) }\dd q \nonumber\\
  &= \int_{0}^\infty e^{-\beta E} \Omega(E) \dd E, \label{eq:Z}
\end{align}
where, without loss of generality we have assumed that $V(q)\ge 0$ on
$\mathcal{D}$.  In terms of the (dimensionless) microcanonical
entropy $S(E)$ and the canonical free energy $G(\beta)$ defined as
\begin{equation}
  \label{eq:entropy}
  S(E) = \log\, \Omega(E) \qquad {\rm and} \qquad G(\beta)  = - \log Z_q(\beta),
\end{equation}
we can write~\eqref{eq:Z} as
\begin{align}
  e^{-G(\beta)}& = \int_0^\infty e^{-\beta E + S(E) } \dd E. \label{eq:Zasympt}
\end{align}
Now suppose that the system size is large enough that the
integral in~\eqref{eq:Zasympt} can be estimated asymptotically by
Laplace's method. If that is the case, then 
\begin{equation}
  \label{eq:9}
  G(\beta) \sim \min_{E\ge 0} \left( \beta E - S(E)\right),
\end{equation}
where $\sim$ means that the ratio of the terms on both sides of the
equation tends to 1 as the system size goes to infinity
(thermodynamic limit). Equation~\eqref{eq:9} states that the free
energy~$G(\beta)$ is asymptotically given by the Legendre-Fenchel
transformation of the entropy $S(E)$. By the involution property of
this transformation, this implies,
\begin{equation}
  \label{eq:9inv}
  S_*(E) \sim \min_{\beta\ge 0} \left( \beta E - G(\beta)\right),
\end{equation}
where $S_*(E)$ is the concave envelope of $S(E)$. This asymptotic relation
can also be written as
\begin{equation}
  \label{eq:12}
  e^{-S_*(E)} \asymp \int_0^\infty e^{-\beta E+ G(\beta)}\dd
  \beta = \int_0^\infty e^{-\beta E}Z_q^{-1}(\beta)\dd \beta,
\end{equation}
where $\asymp$ means that the ratio of the logarithms of the terms on
both sides of the equation tends to 1 as the system size goes to
infinity. As a result
\begin{equation}
  \label{eq:13a}
  \Omega(E) \int_{0}^\infty  e^{-\beta E}Z_q^{-1}(\beta)\dd \beta =
  e^{S(E)} \int_{0}^\infty  e^{-\beta E+G(\beta)}\dd \beta \asymp e^{S(E)-S_*(E)}.
\end{equation}
Comparing with~\eqref{eq:rhocE}, we see that if we set
$\omega(\beta)= Z_q^{-1}(\beta)= e^{G(\beta)}$ we have  
\begin{equation}
  \label{eq:13}
  \bar \rho(E) = \Omega(E) \int_\betamin^\betamax  e^{-\beta E+G(\beta)}\dd \beta
  = e^{S(E)-S_+(E)}, 
\end{equation}
where we have defined
\begin{equation}
  \label{eq:16}
  S_+(E) = -\log \int_\betamin^\betamax  e^{-\beta E+G(\beta)}\dd \beta
  \sim \min_{\beta\in[\betamin,\betamax]} \left( \beta E - G(\beta)\right).
\end{equation}
As a result $\bar \rho(E) \asymp 1$, as desired, provided that
\begin{itemize}
\item [(i)] the system energy $E$ is such that the minimizer
  of~\eqref{eq:9inv} lies in the interval
  $[\beta_{{\rm min}},\beta_{{\rm max}}]$, i.e. $S_+(E) \sim S_*(E)$,
  and
\item [(ii)] $S_*(E)=S(E)$ (i.e. $S(E)$ is concave down).
\end{itemize}
If $V(q)$ is bounded not only from below but also from above,
i.e. $V(q)\le E_{{\rm max}}<\infty$ then the range of possible
system's energies $E$ is $[0, E_{{\rm max}}]$ and we can adjust the
interval $[\beta_{{\rm min}},\beta_{{\rm max}}]$ so that the
minimizer of~\eqref{eq:9inv} always lies in it. If $V(q)$ is unbounded
from above, which is the generic case, this is not possible unless we
let $\betamin \to 0$ (i.e. allow infinitely high temperatures). This
limit breaks ergodicity of the dynamics, and cannot be implemented in
practice. Rather, for unbounded $V(q)$, it is preferable to adjust
$\betamin$ to control the value of $E_{\rm max}$ up to which
$\bar \rho(E)$ is flat. Similarly, regulating $\betamax$ allows
one to keep $\bar \rho(E)$ flat up to some $E_{{\rm min}}>0$ but not
below it.

A more serious problem arises if $S_*(E)\not = S(E)$, i.e. if $S(E)$
is not concave down, since in this case we cannot flatten
$\bar \rho(E)$ in the region where $S(E)$ does not coincide with its
concave envelope. This is the signature that a first-order phase
transition happens. Indeed, a non-concave $S(E)$ implies that the free
energy $G(\beta)$ (which, in contrast to $S(E)$, is always concave
down) is not differentiable at at least one value of~$\beta$:
$G(\beta)$ at each of these values is the transform of a non-concave
branch of $S(E)$. On these branches, ST does not flatten
$\bar \rho(E)$ and as a result it does not \textit{a~priori} provide a
significant acceleration over direct sampling with the original
equations of motion. These observations give an alternative
explanation to a well-known problem of ST in the presence of a
first-order phase transition.

It is also useful to analyze the implications of these results in terms
of the limiting dynamics~\eqref{eq:continuouslim}. If we set
$\omega(\beta)= Z_q^{-1}(\beta)= e^{G(\beta)}$ it is easy to see
from~\eqref{eq:3} that
\begin{equation}
  \label{eq:17}
  \bar V(q) = -\beta^{-1} \log \int_\betamin^\betamax  e^{-\beta_c
    V(q)+G(\beta_c)}\dd \beta_c
  = \beta^{-1} S_+(V(q)), 
\end{equation}
which also implies that $\bar \beta(E) = S'_+(E)$.  As a result,
\eqref{eq:continuouslim} becomes
\begin{equation}
  \begin{split}
    \dot q =& m^{-1} p ~, \\ \dot p =& - \beta^{-1} S'_+(V(q))
    \nabla V(q) - \gamma p + \sqrt{2 \gamma \beta^{-1} m^{-1} } \,
    \eta_p ~, 
    \label{eq:continuouslimwang}
  \end{split}
\end{equation}
The equations used in the Wang-Landau method are similar
to~\eqref{eq:continuouslimwang} but with $S_+(E)$ replaced by $S(E)$.
In other words, these equations are asymptotically equivalent
to~\eqref{eq:continuouslimwang} if $ S_+(E)\sim S(E)$. This is not
surprising since this method also leads to a flat $\bar \rho(E)$. In
other words, ST in the infinite switch limit with
$\omega(\beta)= Z_q^{-1}(\beta)$ is conceptually equivalent to the
Wang-Landau method, although in practice the two methods
differ. Indeed in ST we need to learn $Z_q(\beta)$ to adjust the
weights, whereas in the Wang-Landau method we need to learn its dual
$\Omega(E)$.

\subsection{Adaptive Learning of the Temperature Weights}
\label{sec:learning}

Given any set of weights $\omega(\beta)$ we can in principle learn
$Z_q(\beta)$ (or ratios thereof) using the
estimator~\eqref{eq:Zcestimator}. However we know from the results in
Sec.~\ref{sec:DOS} that this procedure will be inefficient in practice
unless $\omega(\beta) \propto Z^{-1}_q(\beta)$. Here we show how to
adjust $\omega(\beta)$ as we learn $Z_q(\beta)$. To this end we
introduce two quantities: $z(t,\beta_c)$, constructed in such a way
that it converges towards a normalized variant of $Z_q(\beta_c)$; and  $\omega(t,\beta_c)$, giving the current
estimate of the weights.

\begin{enumerate}[leftmargin=.2in]
  \item $z(t,\beta_c)$ with $\beta_c \in [\betamin,\betamax]$ is given by
\begin{equation}
  \label{eq:zq20}
  z(t,\beta_c)  = \frac{1}{t} \int_{0}^{t} \frac{ e^{-\beta_c   V(q(s))} }
  {\int_{\betamin}^{\betamax} \omega(s,\beta_c')
    e^{-\beta_c' V(q(s))} \dd \beta_c' } \dd s.
\end{equation}
\item $\omega(t,\beta_c)$ with $\beta_c \in [\betamin,\betamax]$ is
  the instantaneous estimate of the weights, normalized so that
\begin{equation}
  \label{eq:normaa}
  \int_{\betamin}^{\betamax} \omega(t,\beta_c) \dd \beta_c =1, \qquad \forall t\ge0,
\end{equation}
and satisfying
\begin{equation}
  \tau \dot \omega(t,\beta_c) =  z^{-1}(t,\beta_c) - \lambda(t) \omega(t,\beta),
  \quad {\rm with} \quad \lambda(t) =
  \int_{\betamin}^{\betamax} z^{-1}(t,\beta_c)\ud\beta_c,
  \label{eq:fixedp}
\end{equation}
where $\tau>0$ is a parameter (defining the time-scale with which the
inverse weights are updated in comparison to the evolution of the
physical variables) and $\lambda(t)$ is a renormalizing factor added
to guarantee that the dynamics in~\eqref{eq:fixedp} preserve the
constraint~\eqref{eq:normaa} (the form of this term will become more
transparent when looking at \eqref{eq:zeta1}--\eqref{eq:23}, which are
the time-discretized version of~\eqref{eq:fixedp} considered in
Sec.~\ref{sec:Disc}). Equation \eqref{eq:fixedp} should be solved with the
initial condition $\omega(0,\beta_c) = \omega_0(\beta_c)$, where
$\omega_0(\beta_c)$ is some initial guess for the weights consistent
with~\eqref{eq:normaa}, i.e. such that
$\int_{\betamin}^{\betamax} \omega_0(\beta_c) \dd \beta_c =1$. In
addition, in~\eqref{eq:zq20} $q(s)$, should be obtained by
solving~\eqref{eq:continuouslim} with $\omega(t,\beta) $ substituted
for $\omega(\beta)$, i.e. using
\begin{equation}
  \begin{split}
    \dot q =& m^{-1} p ~, \\ \dot p =& - \beta^{-1} \hat \beta(t,V(q))
    \nabla V - \gamma p + \sqrt{2 \gamma \beta^{-1} m^{-1} } \dot
    \eta_p ~,	\label{eq:continuouslimtdep}
	\end{split}
\end{equation}
where $\hat \beta(t,V(q)) $ is given by
\begin{equation}
  \label{eq:4tdep}
  \hat \beta(t,V(q)) = \frac{\int_{\betamin}^{\betamax} \beta_c
    \omega(t,\beta_c) e^{-\beta_c  V(q)} \dd
    \beta_c }{\int_{\betamin}^{\betamax} \omega(t,\beta_c) e^{-\beta_c  V(q)} \dd
    \beta_c}.
\end{equation}
\end{enumerate}

To understand these equations, suppose first that $\tau=\infty$. In
this case, $\omega(t,\beta_c)$ is not evolving, i.e.
$\omega(t,\beta_c) = \omega_0(\beta_c)$ for all $t\ge 0$. It is then
clear that~\eqref{eq:continuouslimtdep} reduces to
\eqref{eq:continuouslim} and \eqref{eq:zq20} to the estimator at the
right hand side of~\eqref{eq:Zcestimator} as $t \to \infty$. In other
words, when $\tau=\infty$ we have
\begin{equation}
  \lim_{t\to \infty} z(t,\beta_c) = \frac{Z_q(\beta_c)}{ 
    \int_{\betamin}^{\betamax} Z_q(\beta'_c) \omega_0(\beta'_c)\ud\beta'_c},
    \label{eq:Zest}
\end{equation}
i.e. for any $\beta_c$, $\beta_c'$,
\begin{equation}
  \label{eq:18}
  \lim_{t\to \infty} \frac{z(t,\beta_c)}{z(t,\beta_c')}= \frac{Z_q(\beta_c)}{ 
    Z_q(\beta_c') }.
\end{equation}

When $\tau <\infty$, $\omega(t,\beta_c)$ is evolving and we need to
consider the fixed point(s) of this equation. Suppose that at least
one such a fixed point exists and denote it by $\omega_\infty(\beta)$. Since
this fixed point must satisfy
$\int_{\betamin}^{\betamax} \omega_\infty(\beta_c)\ud\beta_c =1$, it is easy
to see from~\eqref{eq:fixedp} that it must be given by
\begin{equation}
  \label{eq:21}
  \omega_\infty(\beta_c) = \frac{z_\infty^{-1}(\beta_c)}
  { \int_{\betamin}^{\betamax} z_\infty^{-1}(\beta'_c)\ud\beta'_c}, 
\end{equation}
where $z_\infty(\beta_c) = \lim_{t\to\infty} z(t,\beta_c)$ which,
from~\eqref{eq:zq20} is given by (replacing again $\omega(t,\beta_c) $ by
$ \omega_\infty(\beta_c)$):
\begin{equation}
  \label{eq:20}
  \begin{aligned}
    z_\infty(\beta_c) & = \lim_{T\to\infty} \frac{1}{T} \int_{0}^{T}
    \frac{ e^{-\beta_c V(q(t))} } {\int_{\betamin}^{\betamax}
      \omega_\infty(\beta'_c) e^{-\beta'_c V(q(t))} \dd \beta'_c } \dd
    t\\
    & = \int_{\mathcal{D}\times \mathbb{R}^d} \frac{ e^{-\beta_c V(q)}
    } {\int_{\betamin}^{\betamax} \omega_\infty(\beta'_c) e^{-\beta'_c
        V(q)} \dd \beta'_c } \bar \rho(q,p) \dd q \dd p
  \end{aligned}
\end{equation}
where we used ergodicity and $\bar \rho(q,p)$ is the equilibrium
density of~\eqref{eq:continuouslim} when the weights
$\omega_\infty(\beta_c)$ are used to calculate $\bar \beta(V(q))$ in
this equation. It can be checked directly that~\eqref{eq:21}
and~\eqref{eq:20} admit as a solution
\begin{equation}
  \label{eq:22}
  \omega_\infty(\beta_c) = \frac{Z^{-1}_q(\beta_c)}
  {\int_{\betamin}^{\betamax} Z_q^{-1}(\beta'_c)\ud\beta'_c}.
\end{equation}
and that 
\begin{equation}
  \label{eq:18b}
  \frac{z_\infty(\beta_c)}{z_\infty(\beta_c')}= \frac{Z_q(\beta_c)}{ 
    Z_q(\beta_c') },
\end{equation}
for any $\beta_c$, $\beta_c'$.

The argument above implies the existence of a fixed point
of~\eqref{eq:fixedp} that satisfies~\eqref{eq:21}, i.e. that can be
used as optimal weight $\omega(\beta_c) \propto
Z_q^{-1}(\beta_c)$. This argument does not indicate the conditions
under which this fixed point is stable, nor the size of its basin of
attraction. General considerations based on averaging theorems for
systems with multiple timescales suggest that this fixed point should
be stable from any initial condition in the limit $\tau\to\infty$,
provided that we look at the evolution of $\omega(t,\beta)$ on its
natural $O(\tau)$ timescale. In Sec.~\ref{sec:NumExp} we will verify
using numerical examples that the fixed point~\eqref{eq:22} is also
reached for moderate values of $\tau$.

%
\section{Implementation details of the ISST algorithm}
\label{sec:Disc}
%

Let us now discuss the practical aspects of the ISST algorithm.

For the purpose of discretizing the limiting
equation~(\ref{eq:continuouslim}), we suggest to use the second order
``BAOAB'' Langevin scheme\cite{Leimkuhler}, with equations
\begin{equation}
   \label{eq:Dis-isst}
  \begin{aligned}
    p_{n+1/2} &= p_n - \tfrac{1}{2} \Delta t\beta^{-1}\bar{\beta}( V(q_n) )\nabla V(q_n),\\
    q_{n+1/2} &= q_n + \tfrac{1}{2} \Delta t m^{-1} p_{n+1/2},\\
    \hat{p}_{n+1/2} &= e^{-\Delta t\gamma} p_{n+1/2} +
    [\beta^{-1}(1-e^{-2 \gamma \Delta t})m]^{1/2} \eta_n,\\
    q_{n+1} &= q_{n+1/2} + \tfrac{1}{2} \Delta t m^{-1} \hat{p}_{n+1/2},\\
    p_{n+1} &= \hat{p}_{n+1/2} - \tfrac{1}{2}\Delta t
    \beta^{-1}\bar{\beta}( V(q_{n+1}) ) \nabla V(q_{n+1}),
  \end{aligned}
\end{equation}
where $(q_n,p_n)$ are the time-discretized approximations of
$(q(n\Delta t), p(n\Delta t))$, $\Delta t$ is the timestep, and
$\eta_n\sim \mathcal{N}(0,1)$.  This method is known to have low
configurational sampling bias in comparison with other Langevin MD
schemes \cite{LeMa2013b}.

In order to make the scheme above explicit, one needs to
estimate~\eqref{eq:4}, i.e. provide a scheme to evaluate
$\bar{\beta}( V(q_n) )$ given the value of the potential
$V(q_n)$. This involves two issues: the first is how to estimate the
$1$--dimensional integrals in \eqref{eq:4} given the weights $\omega(t,\beta_c)$;
the second is how to update the weights by discretizing the equations
given in Sec.~\ref{sec:learning}.

Regarding the first issue, any quadrature (numerical
integration) method can in principle be used.  However, since this
quadrature rule is part of an iterative `learning' strategy in which
statistics are accumulated on-the-fly to update the weights, it is
desirable to use a fixed set of nodes or grid points
$\{\beta_i \}_{1\leq i \leq M}$, so that the corresponding samples
collected at earlier stages remain relevant as the system is updated.

For a fixed number of nodes, the optimal choice of quadrature rule on
a given interval $[\betamin,\betamax]$ in terms of accuracy is derived
by placing the nodes at the roots of a suitably adjusted Legendre
polynomial (Gauss-Legendre quadrature). Let the quadrature weight for
node $i$ be $B_i$ and replace $\bar\beta(V(q_n))$
in~\eqref{eq:continuouslim} by,
\begin{equation}
  \hat\beta(V(q_n)) = \frac{ \sum_{i=0}^{M} B_i \beta_{i}
    \omega_{i,n}
    e^{ -\beta_i V(q_n)}}
  {\sum_{i=0}^{M} B_i \omega_{i,n} e^{ -\beta_i V(q_n)}}.
  \label{eq:BetaDisc}
\end{equation} 
where $\omega_{i,n}$ is the current estimate of the weight at node
$i$.

To obtain $\omega_{i,n}$ we use the following discrete recurrence
relation consistent with~\eqref{eq:fixedp}. Given $\omega_{i,n}$, we
find $\omega_{i,n+1}$ via,
\begin{equation}
  \label{eq:zeta1}
  \omega_{i,n+1} = \frac{\omega_{i,\star}}{ \sum_{j=1}^{M} B_j \omega_{j,\star} }~, 
\end{equation}
in which
\begin{equation}
  \label{eq:zeta2}
  \omega_{i,\star} = \left( 1 - \tau^{-1} \Delta t \right) \omega_{i,n}
  +  \tau^{-1} \Delta t  \, z^{-1}_{i,n},
\end{equation}
with
\begin{equation}
  \label{eq:23}
  z_{i,n} = \frac1n \sum_{m=1}^n \frac{e^{-\beta_i V(q_m)}}{ 
    \sum_{j=1}^{M} B_j \omega_{j,m}  e^{-\beta_j V(q_m)} }.
\end{equation}
Here, $\tau>0$ is the time-scaling parameter
introduced in Sec.~\ref{sec:learning}. It can be checked that this
recurrence relation preserves the constraint that for all~$n \geq 0$,
\begin{equation}
	\label{eq:24}
	\sum_{j=1}^{M} B_j \omega_{j,n}   = 1~,
\end{equation}
and that \eqref{eq:zeta1} and \eqref{eq:zeta2} are consistent
with~\eqref{eq:normaa}. Note also that \eqref{eq:23} can be
written in terms of the following iteration rule
\begin{equation}
  \label{eq:23it}
  z_{i,n} = \frac1n \frac{e^{-\beta_i V(q_n)}}{ 
    \sum_{j=1}^{M} B_j \omega_{j,n}  e^{-\beta_j V(q_n)} } + \frac{n-1}{n}z_{i,n-1}.
\end{equation}
and that it gives a running estimate of the ratio in \eqref{eq:Zest}.

The discussion above makes apparent that ISST requires very few
adjusting parameters besides the one already used in a vanilla MD code
(i.e. the parameters like $\Delta t$ in~\eqref{eq:Dis-isst}):
Basically, the user is required to choose the temperature range
$[\betamin,\betamax]$ and the scaling parameter $\tau$. Other
parameters, like the number and positions of the nodes for the
temperature or the quadrature rule to be used, can be adjusted using
standard practices in numerical quadrature, and they do not affect the
overal efficiency of the method.

%
\section{Numerical Experiments}
\label{sec:NumExp}
%
In order to evaluate the performance of the discretization method
described in the previous section we now present results from several
numerical experiments on three test systems: the $d$--dimensional
harmonic oscillator; the continuous Curie-Weiss model, a mean field
version of the Ising model which displays a second-order phase
transition in temperature; and the Alanine-12 molecule in vacuum,
which displays conformation changes. On these examples we investigate
(i) the influence of the number of quadrature points, $M$, used in the
ISST algorithm when the weights $\omega_i$ are known; (ii) the
convergence of \eqref{eq:23}, both when the weights are fixed to their
initial (and non-optimal) value ($\tau\to\infty$ limit) and when these
weights are adjusted towards their optimal values; and (iii) the
effect of the choice of $\tau$ on the convergence of the weights
$\omega_i$, estimated using \eqref{eq:zeta1}-\eqref{eq:23}. We also
compare the efficiency of ISST and standard ST on the last two
examples.

\subsection{Harmonic Oscillator}
\label{sec:DHarm}

\begin{figure}[!ht]
	\centering
	\includegraphics{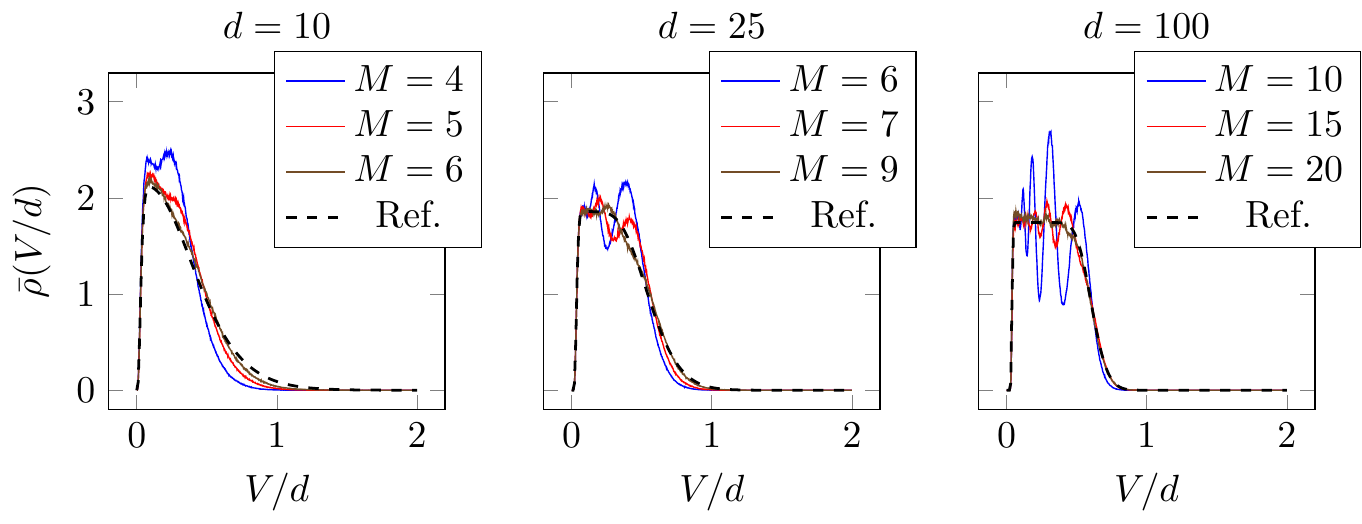}
	\caption{Behavior of $\bar{\rho}(E)$ with $E=V/d$, for the
          harmonic oscillator. The results for different numbers of
          temperatures $M$ were obtained by recording the energy of an
          ISST trajectory of length $N=10^7$ steps using a
          histogram. The ISST weights are given by
          \eqref{eq:discHamWeight}.}
	\label{fig:NumGpnts}
\end{figure}

Consider a $d$-dimensional harmonic oscillator given by the
quadratic potential in $\mathbb{R}^{d}$,
\begin{equation}
\label{eq:HamPot}
	V(q) = \frac{1}{2} \sum_{j=1}^{d} \lambda_j q_j^2 ,
\end{equation}
where $\{\lambda_j\}_{j=1}^d$ is a set of positive constants. The
partition function $Z_q( \beta )$ can be written explicitly as,
\begin{equation}
	\label{eq:ZHam}
	Z_q( \beta ) = A
	\beta^{-d/2} \qquad {\rm with} \quad 
	A = (2 \pi)^{d/2} \prod_{j=1}^d \lambda_j^{-1/2}.
\end{equation}
The goal is to perform simulations using \eqref{eq:Dis-isst} and
\eqref{eq:BetaDisc} with some yet to be determined
weights~$\omega_i$. As we showed in Sec.~\ref{sec:DOS} the asymptotic
optimal weight is $\omega(\beta) \propto Z_q^{-1}(\beta)$, which
implies that $\omega_i \propto Z_q^{-1}(\beta_i)$. This leads to a
log-asymptotically flat energy in \eqref{eq:13} i.e
$\bar{\rho}(E) \asymp 1$.

Using \eqref{eq:ZHam} we can write the density of states, for the
potential given by \eqref{eq:HamPot} as,
\begin{equation}
	\Omega(E) = \frac{A  E^{1-d/2}}{\Gamma(\tfrac{d}2)}.
\end{equation}
Additionally, since $\Omega^{-1}(E) $ is completely monotonic for
$d>2$, the Hausdorff-Bernstein-Widder-theorem guarantees the existence
of a measure $\mu(\beta_c)$ such that,
\begin{equation}
	\label{eq:10}
	\Omega^{-1}(E) = \int_0^\infty e^{-\beta_c E} \dd\mu(\beta_c).
\end{equation}
It is straight forward to verify that this measure is
\begin{equation}
d\mu(\beta_c) = D^{-1} \beta_c^{ d/2-2 } d\beta_c, 
\qquad {\rm with} \quad 
D = \frac {\Gamma(\tfrac{d}2-1) \, A}{\Gamma(\tfrac{d}2)}.
\end{equation}
With the knowledge of \eqref{eq:10}, it is easy to see that \eqref{eq:13} requires that $\bar{\rho}(E) = 1$ as $d\to\infty$ if $\omega(\beta_c) \propto \beta_c^{ d/2-2 }$. This suggests that one could use,
\begin{equation}
\omega(\beta_c) \propto
\left \{  \begin{array}{lcl}
\beta_c^{ d/2 -2},  & \qquad & \beta_c \in[\betamin,\betamax]\\
0, & & {\rm else}
\end{array} \right .
\label{eq:15}
\end{equation}
The arbitrary constant of proportionality is determined so as to satisfy \eqref{eq:24}, such that the explicit form of the $M$ weights in \eqref{eq:BetaDisc} is defined by,
\begin{equation}
\label{eq:discHamWeight}
\omega_i = \beta_i^{d/2-2}\left( \sum_{j=1}^M B_j \beta_j^{d/2-2} \right)^{-1}.
\end{equation}

In Figure \ref{fig:NumGpnts} we show results for $d = 10,~25$ and
$100$, sampled using \eqref{eq:Dis-isst}, \eqref{eq:BetaDisc} and
\eqref{eq:discHamWeight} with a total trajectory length of $N=10^7$
with $\Delta t = 0.1$ and $\betamin=0.8$ and $\betamax=12.5$. Each
panel shows the convergence to the dashed reference \eqref{eq:13}
found using quadrature. We conduct experiments for three values of the
dimension $d$, in which we vary the number of quadrature points (or
reciprocal temperatures) $M$. The figure clearly illustrates the
importance of choosing an appropriate number of points $M$, such that
the observable of interest has a satisfactory support. Also note the
dependence of $M$ on the dimension $d$, which is an entropic effect
resulting from the dependence of the potential \eqref{eq:HamPot} on
$d$.

\subsubsection{Adaptive Weight Learning for the Harmonic Oscillator}
\label{sec:AWLA}
\begin{figure}[!ht]
	\centering
	\includegraphics{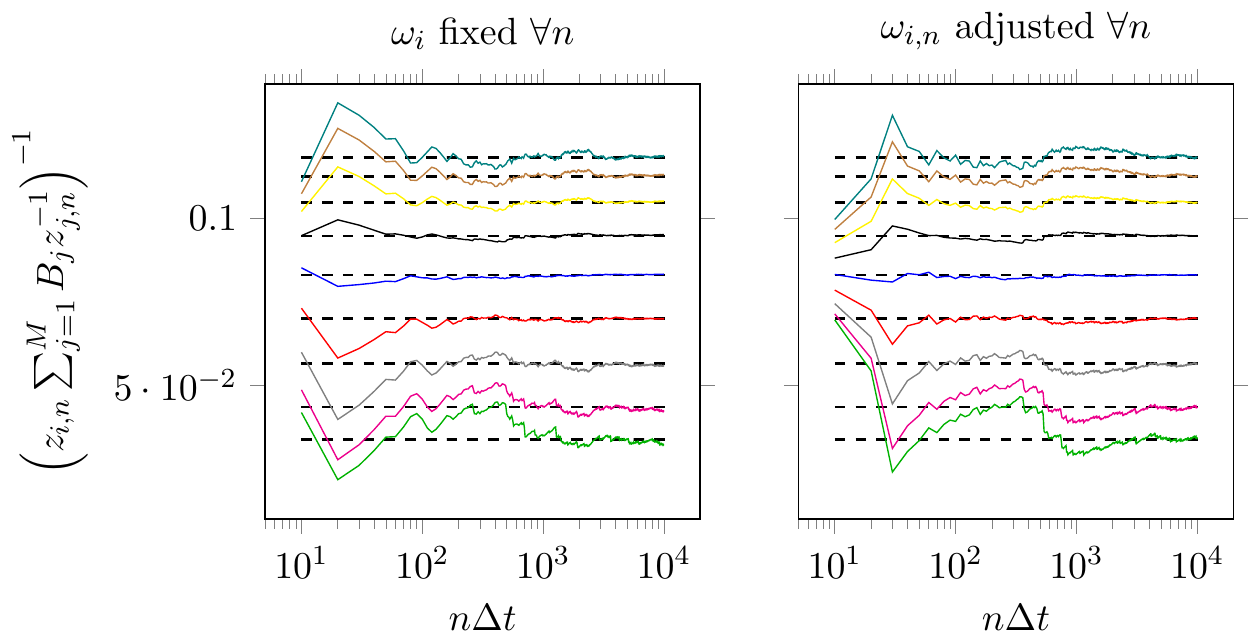}
	\caption{Reciprocal of \eqref{eq:fixedp2} learned using $M=10$
          temperatures between $\betamax=12.5$ and $\betamin=0.8$ for
          \eqref{eq:HamPot} with $d=1$ and $\Delta t = 0.01$ with
          $\tau=1$ in \eqref{eq:zeta2}. The black dashed lines show
          the asymptotic long-term average \eqref{eq:HarmAsym}. In the
          left panel we keep the weight fixed for all simulation time
          and in the right panel we update the weights at every
          timestep.}
	\label{fig:QPlots}
\end{figure}
In this section we check the convergence in time of the following
quantity,
\begin{equation}
  \label{eq:fixedp2}
  \qquad z_{i,n} \sum_{j=1}^{M} B_j z^{-1}_{j,n},
\end{equation}
with $z_{i,n}$ given by~\eqref{eq:23}.  This is a normalized version
of the partition function whose inverse gives the optimal
weight~\eqref{eq:21}.

We perform a comparative experiment between two variants of the
estimate \eqref{eq:fixedp2}. First we initialize the weights at
$\omega_i \propto 1$, normalize according to \eqref{eq:24} and fix
these weights for a complete ISST simulation. Secondly, we instead
initialize $\omega_{i,0}\propto 1$ and normalize according to
\eqref{eq:24}, and adjust the weights at every timestep as described
in Sec.~\ref{sec:Disc}.

In Figure \ref{fig:QPlots} we show the results of these experiments
using \eqref{eq:HamPot} with $d=1$ and $M=10$ reciprocal temperatures
between $\betamin=0.8$ and $\betamax=12.5$. In the left panel we
present the results of the first experiment described above, in which
we fix the weights $\omega_i$ for all simulation time--as indicated by
the title. To the right, we show the second experiment in which we
adjust the weights at every timestep via \eqref{eq:zeta1},
\eqref{eq:zeta2} and \eqref{eq:24}.

Both panels in Figure \ref{fig:QPlots} show the reciprocal of
\eqref{eq:fixedp2} for all the $M$ temperatures in color, whereas in
dashed black we show the time asymptotic behaviour. The
time-asymptotic limit as $n\to\infty$ in \eqref{eq:fixedp2} is:
\begin{equation}
\label{eq:HarmAsym}
	\beta_i^{1/2}\left( \sum_{j=1}^{10} B_j \beta^{1/2}_{j}\right)^{-1}.
\end{equation}

It is clear from Figure \ref{fig:QPlots} that it is possible to learn
ratios of the partition functions for a modest number of timesteps
$n$, regardless of the value of $\omega_i$. In practice one does not
wish to fix the weights at some non-optimal value, as was done
initially in this section, as this will most likely impede the
sampling efficiency of the algorithm. Instead it is preferable to make
use of the second approach, where one adjusts the weights continuously
towards some optimum, as the simulation progresses.

\subsubsection{Convergence of the Temperature Weights}
\label{sec:CoTW}

The combination of the ISST Langevin scheme \eqref{eq:Dis-isst} and
the adaptive weight-learning \eqref{eq:zeta1} results in a powerful,
simple-to-implement sampling algorithm. In Sec.~\ref{sec:learning} we
introduced a timescale parameter $\tau$ which adjusts the rate of
weight learning in relation to the timestep in the ISST scheme. This
section aims to explore the choice of this parameter and its effect on
the convergence of the weights.
\begin{figure}[!ht]
	\centering
	\includegraphics{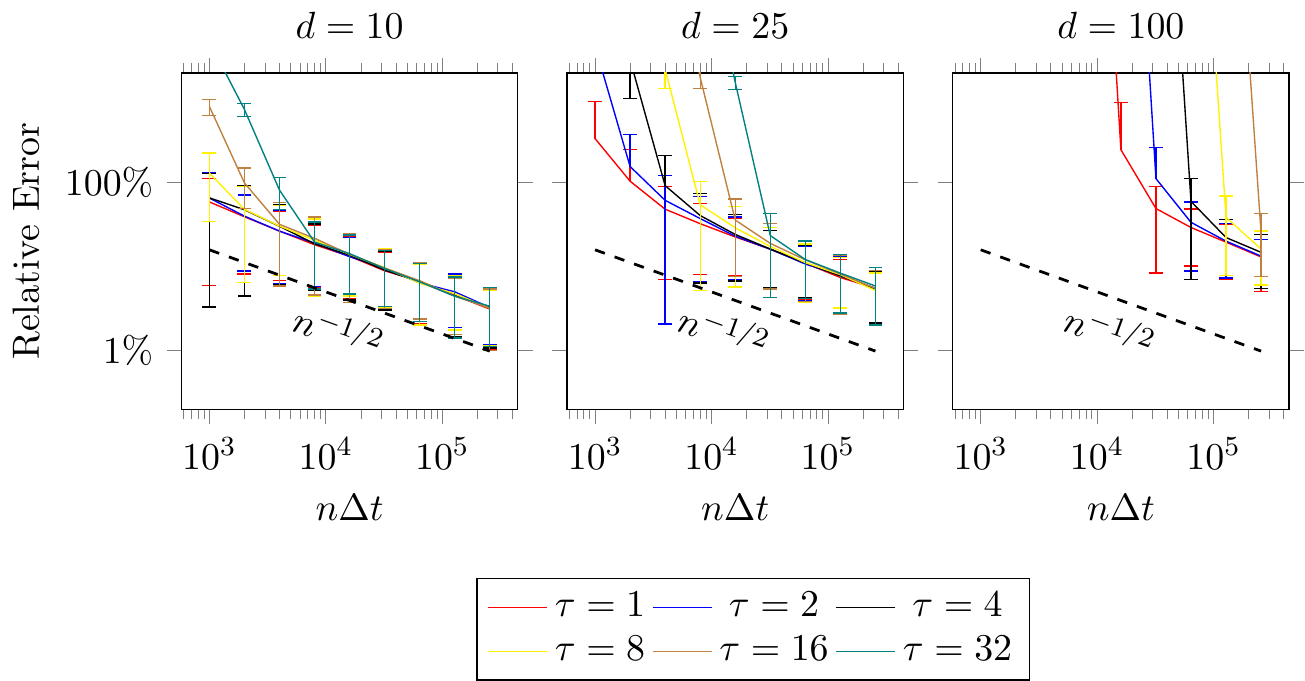}
	\caption{Convergence rate for a $d$-dimensional harmonic
          oscillator using $M=10$ temperatures between
          $\betamin = 0.08$ and $\betamax = 12.5$ with $\Delta t=0.1$
          and a wide range of $\tau$. As $\tau$ is increased the
          dynamics of the weight-learning algorithm slows down, which
          consequently slows down the convergence of the weights. The
          convergence rate $n^{-1/2}$ is the standard error decay of
          Monte Carlo averages. Each data point was calculated by
          averaging over 200 independent ISST trajectories and the
          error bars are the standard deviations associated with these
          averages.}
	\label{fig:learnconv}
\end{figure}

We define the relative error as
\begin{equation}
  \label{eq:RelErr}
  {\rm Rel. Error} = \sup_{M \geq i > 0} \frac{\vert \omega_{i,N}
    - \omega_{i,\infty} \vert}{\vert \omega_{i, \infty} \vert}
  \qquad {\rm with} \qquad
  \omega_{i,\infty} ={\beta}_i^{d/2} \left( \sum_{j=1}^{M} B_j {\beta}_j^{d/2}\right)^{-1},
\end{equation}
where $N$ refers to the last timestep of the simulation. We use
\eqref{eq:RelErr} as a metric for the accuracy of the approximations
from \eqref{eq:zeta1}, i.e $\omega_{i,n}$ for $0\leq n \leq N$.
Working with $M=10$ temperatures between $\betamin = 0.08$ and
$\betamax = 12.5$ we perform experiments using \eqref{eq:Dis-isst}
with $\Delta t=0.1$ varying $\tau$ in~\eqref{eq:zeta2}. The results of
this experiment with initial condition $\omega_{i,0}\propto1$ for all
$i$, are shown in Figure \ref{fig:learnconv}. In
Sec.~\ref{sec:learning} we indicated that the fixed point of the
learning scheme should be stable as $\tau \to \infty$ and we now
observe that, at least in this example, it is stable even for moderate
values of $\tau$. In fact it appears that there is no advantage of
using $\tau$ large and we see that its only effect, in the toy model,
is to slow the convergence to the fixed point.  In more complicated
systems the choice of $\tau$ will be more critical.

The previous section implied that the adjustment scheme
\eqref{eq:zeta1} for the weights $\omega_{i,n}$, should be dependent
on the approximation of the ratio of partition functions
$z_{i,n}$. Figure \ref{fig:learnconv} makes it clear that the
estimation of $z_{i,n}$ dominates the error when estimating the
weights $\omega_{i,n}$. Consequently the observed $n^{-1/2}$
convergence with timestep is a result of this Monte Carlo
averaging. Accuracy in $\omega_{i,n}$ can therefore only be gained by
extending simulation time.

\subsection{Curie-Weiss Magnet}

We next consider a continuous version of the Curie-Weiss magnet,
i.e. the mean field Ising model with $K$ spins and potential
\begin{equation}
  \label{eq:CWpot}
  V_K(\theta_1, \ldots,
    \theta_K;b) = - \frac{1}{2K} \left( \sum_{i=1}^{K} \cos\theta_i
  \right)^2 - b \sum_{i=1}^{K} \cos \theta_i .
\end{equation}
where $b\in \mathbb{R}$ is the intensity of the applied field.  The Gibbs
(canonical) density for this model is,
\begin{equation}
  \label{eq:gibbstheta}
  \varrho_K(\theta_1, \ldots,
    \theta_K,\beta,b) =
  \mathcal{Z}^{-1}_K(\beta,b) \exp\left[-\beta V_K(\theta_1, \ldots,
    \theta_K;b)\right],
\end{equation}
where,
\begin{equation}
  \label{eq:parttheta}
  \mathcal{Z}_K(\beta,b) = 
  \int_{[-\pi,\pi]^{K}} \exp\left[-\beta V_K(\theta_1, \ldots,
    \theta_K;b)\right]
  d\theta_1 \ldots \theta_K.
\end{equation}
This system has similar thermodynamics properties as the standard
Curie-Weiss magnet with discrete spins, but it is amenable to
simulation by Langevin dynamics since the angles $\theta_i$ vary
continuously. That is, we can simulate it in the context of ST in the
infinite switch limit using~\eqref{eq:continuouslim} with $(\theta_1,
\ldots, \theta_K)$ representing the role of $q$.

\subsubsection{Thermodynamic properties and phase transition diagram}
\label{sec:thermo}

Like the standard Curie-Weiss magnet, the system with
potential~\eqref{eq:CWpot} displays phase-transitions when $\beta$ is
varied with $b=0$ fixed and when $b$ is varied with $\beta$ fixed
above a critical value. To see why, and also introduce a quantity that
we wil monitor in our numerical experiments, let us marginalize the
Gibbs density~\eqref{eq:gibbstheta} on the average magnetization $m$
defined as
\begin{equation}
  \label{eq:19}
  m = \frac1K\sum_{i=1}^{K} \cos \theta_i.
\end{equation}
This marginalized density is given by
\begin{equation}
  \label{eq:25}
  \rho_K(m,\beta,b)  = \int_{[-\pi,\pi]^{K}} \varrho_K(\theta,\beta,b)
  \delta \left( m - \sum_{i=1}^{K} \cos \theta_i
  \right)
  d\theta_1 \ldots \theta_K .
\end{equation}
A simple calculation shows that
\begin{equation}
  \label{eq:gibbsm}
  \rho_K(m,\beta,b) = Z^{-1}_K(\beta,b) \exp\left[-\beta K F_K(m;\beta,b)\right],
\end{equation}
where $Z_K(\beta,b) = \int_{-1}^{1} e^{-\beta K F_K(m;\beta,b) }dm$
and we introduced the (scaled) free energy $F_K(m;\beta,b)$ (not to be
confused with the free energy introduced in~\eqref{eq:entropy}, here
given by $G(\beta) = - K^{-1} \log Z_K(\beta,b)$) defined as
\begin{equation}
  \label{eq:33}
  F_K(m;\beta,b) = V(m;b)- \beta^{-1} S_K(m)
\end{equation}
with potential term
\begin{equation}
  \label{eq:CWpot2}
  V(m;b) = -\tfrac12 m^2 - b m, 
\end{equation}
and entropic term 
\begin{equation}
  \label{eq:cwent}
  S_K(m) = K^{-1} \log 
  \int_{[-\pi,\pi]^{K}} \delta \left( m - \sum_{i=1}^{K} \cos \theta_i
  \right)
  d\theta_1 \ldots \theta_K .
\end{equation}

The marginalized density~\eqref{eq:gibbsm} and the free
energy~\eqref{eq:33} can be used to analyze the properties of the
system in thermodynamic limit when $K\to\infty$ and map out its phase
transition diagram in this limit. In particular, we show next that
$F_K(m;\beta,m)$ has a limit as $K\to\infty$ that has a single minimum
at high temperature, but two minima at low temperature. Since
$F_K(m;\beta,m)$ is scaled by $K$ in~\eqref{eq:gibbsm}, this implies
that density can become bimodal at low temperature, indicative of the
presence of two strongly metastable states separated by a free energy barrier
whose height is proportional to $K$.  

The limiting free energy $F(m;\beta,b)$ is defined as
\begin{equation}
  \begin{aligned}
    F(m;\beta,b) &= \lim_{K\to \infty} F_K(m;\beta,b)\\
	&= -\tfrac12 m^2 - b m -\beta^{-1} \lim_{K \to \infty}S_K(m).
\end{aligned}
\label{eq:FreeEnergy}
\end{equation}
\begin{figure}[!ht]
	\centering
	\includegraphics[width=3in]{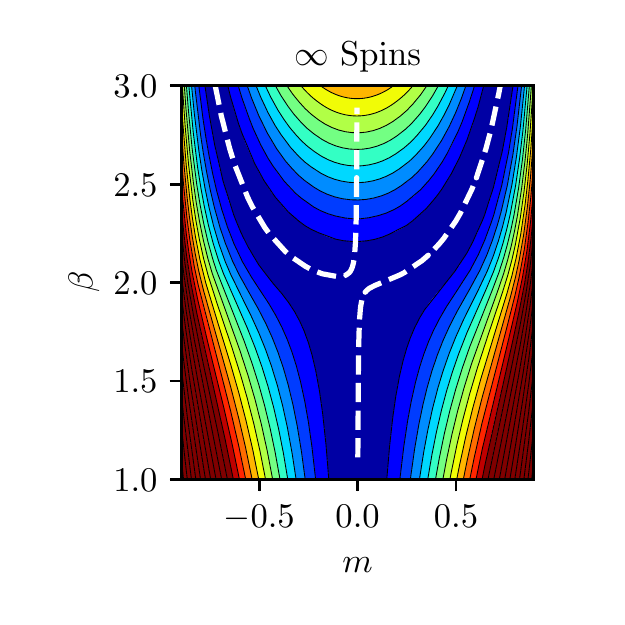}
	\caption{Limiting free energy $F(m;\beta,b)$ as $K\to\infty$ for $(m,\beta)$
          using $b=0.001$. The averaged magnetization where
          $F'(m;\beta,b)=0$ is shown in white and the contour is the
          free energy surface \eqref{eq:FreeEnergy}.}
	\label{fig:LimitPhaseTrans}
\end{figure}
To calculate the limit of the third (entropic) term, let us define
$H(\lambda)$ via the Laplace transform of \eqref{eq:cwent} through
\begin{equation}
\begin{split}
  e^{ -K H(\lambda)} & = \int_{-1}^1 e^{-K \lambda m + K S_K(m)} dm\\
  &= \int_{[-\pi,\pi]^K} e^{-\lambda \sum_{i=1}^{K} \cos(\theta_i)}
  d\theta_1 \ldots \theta_K
  \\
  &= \prod_{i=1}^{K} \int_{-\pi}^{\pi} e^{-\lambda \cos \theta_i} d
  \theta_i \\&= \left( 2 \pi I_0(\lambda) \right)^K,
\end{split}
\end{equation}
where $I_0(\lambda)$ is a modified Bessel function.  In the large $K$
limit, $S(m)$ can be calculated from $H(\lambda)$ by Legendre
transform
\begin{equation}
\begin{split}
  S(m) = \lim_{K\to\infty} S_K(m) =& \min_{\lambda} \left\{ \lambda m
    - H(\lambda) \right\} \\=& \min_{\lambda} \left\{ \lambda m + \log
    I_0(\lambda) \right\} + \log (2\pi).
\label{eq:105}
\end{split}
\end{equation}
The minimizer $\lambda(m)$ of~\eqref{eq:105} satisfies
\begin{equation}
  \label{eq:26}
  m  =  -\frac{I_1(\lambda(m))}{I_0(\lambda(m))}
\end{equation}
which, upon inversion, offers a way to parametrically represent $S(m)$
using
\begin{equation}
  \label{eq:27}
  \begin{aligned}
    S(m(\lambda)) = \lambda m(\lambda) + \log I_0(\lambda) + \log
    (2\pi), \qquad m(\lambda) = -\frac{I_1(\lambda)}{I_0(\lambda)},
    \qquad \lambda \in \mathbb{R}.
  \end{aligned}
\end{equation}
Similarly we can represent $F(m;\beta,b)$ as 
\begin{equation}
  \label{eq:28}
  F(m(\lambda);\beta,b)= -\tfrac12 m^2(\lambda) - b m(\lambda)
  -\beta^{-1} S(m(\lambda)), \qquad m(\lambda) =
  -\frac{I_1(\lambda)}{I_0(\lambda)} \qquad \lambda \in \mathbb{R}.
\end{equation}
In Figure \ref{fig:LimitPhaseTrans} we show a contour plot of
\eqref{eq:FreeEnergy} as a function of $m$ and $\beta$ for fixed $b$
obtained using this representation. Also shown is the location of the
minima of $F(m;\beta,b)$ in the $(m,\beta)$ plane at $b$ fixed. These
minima can also be expressed parametrically. Indeed,~\eqref{eq:27}
implies that
\begin{equation}
  \label{eq:29}
  S'(m(\lambda)) = \lambda
\end{equation}
which if we use it in $F'(m;\beta,b)=0$ to locate the minima of the
free energy in the $(m,\beta)$ plane indicates that they can be
expressed parametrically  as
\begin{equation}
  \beta(\lambda) = \frac{\lambda}{m(\lambda)+b}, \qquad
  m(\lambda) = -\frac{I_1(\lambda)}{I_0(\lambda)} \qquad \lambda \in \mathbb{R}.
\label{eq:ParamBiFu}
\end{equation}
The corresponding path gives the averaged magnetization as a function
of $\beta$ and is shown as a dashed line in
Fig.~\ref{fig:LimitPhaseTrans} and was plotted using these formulae with
$b=0.001$.  For values of $\beta$ less than 2, the free energy is a
single-well, and the averaged magnetization is approximately zero. For
values of $\beta$ above 2, the free energy becomes a double-well, and two
metastable states with nonzero magnetization emerge. 

If we consider the case $b=0$, then by symmetry, $m=0$ is a critical
point of $F(m,\beta,b=0)$ for all values of $\beta$, i.e.
$F'(0,\beta,b=0) = 0$. By differentiating~\eqref{eq:29} in $\lambda$
using the chain rule, we deduce that
 \begin{equation}
   \label{eq:30}
   S''(m(\lambda)) = 1/m'(\lambda)
 \end{equation}
 which, if we evaluate it at $\lambda=0$ using $m(\lambda=0)=0$ as
 well as $m'(\lambda=0) = -\frac12$ which follows from $m(\lambda) =
 -I_1(\lambda)/I_0(\lambda)$, indicates that
 \begin{equation}
   \label{eq:31}
   S''(0) = -2.
 \end{equation}
 As a result
 \begin{equation}
   \label{eq:32}
   F''(0;\beta, b=0) = -1+ 2 \beta^{-1}
 \end{equation}
 which means that $m=0$ is a stable critical point of
 $F(m;\beta, b=0) $ for $\beta< \beta_c =2$, and an unstable critical
 point for $\beta> \beta_c =2$, with a phase transition occurring at
 $\beta_c = 2$. A similar calculation can be performed when $b\not=0$,
 but it is more involved since $m=0$ is not a critical point in this
 case (hence we need to solve \eqref{eq:ParamBiFu} numerically in
 $\lambda$ to express the critical $m$ as a function of $\beta$): in
 this case, the location of the global minimum of $F(m;\beta,b)$
 varies continuously, so strictly speaking there is no phase
 transition. 

 It should be stressed that the phase transition observed when $\beta$
 is varied at $b=0$ fixed is second order, i.e.
 $G(\beta) = - K^{-1} \log Z_K(\beta,b=0)$ is continuous with a
 continuous first order derivative in $\beta$ at $\beta= \beta_c=2$,
 but discontinuous in its second order derivative at that point. As a
 result the phase transition observed in the model above does not
 lead to difficulties of the kind discussed in Sec.~\ref{sec:DOS}: in
 particular it can be checked by direct calculation that
 $S_*(E) = S(E)$ (i.e. the entropy $S(E)$ is concave down).

%
%
%

\subsubsection{Sampling near or at the Phase Transition}

In this section we use \eqref{eq:Dis-isst} with weights adjusted as in
\eqref{eq:zeta1} to sample \eqref{eq:gibbsm} over a range of
temperatures, from high when \eqref{eq:gibbsm} is mono-modal to low
when it is bi-modal. This is challenging for standard sampling methods
because, as indicated by the results in Sec.~\ref{sec:thermo}, at low
temperature the system has two metastable states separated by an
energy barrier whose height scales linearly with $K$ as $K$
increases.
\begin{figure}[!ht]
	\centering
	\includegraphics{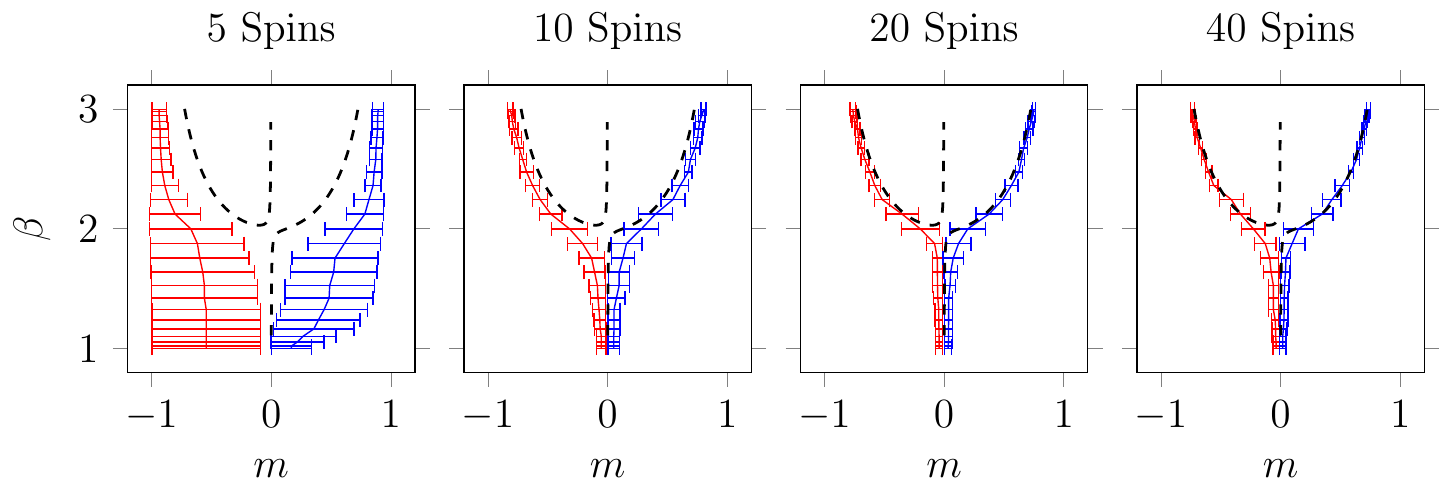}
	\caption{Minimum of the free energy vs the number of spins, in
          comparison to the theoretical minimum for $K\to\infty$ shown
          as dashed black. The different colour lines show the minimum
          in the upper and lower half, respectively. Each data point
          was calculated by averaging over 20 ISST simulations of
          length $N=10^5$ with timestep $\Delta t = 0.1$ and
          $M=25$. The minimum of the magnetization $m$ was found by
          collecting points from the trajectories in a histogram with
          200 bins, the minimum was then found in the upper
          (respectively lower )100 bins and are shown in red (and
          blue).}
	\label{fig:PhaseTrans}
\end{figure}

The results of our experiments using varying numbers of spins are
presented in the four panels of Figure \ref{fig:PhaseTrans}. The
minimum of the sampled free energy in the lower half of the
magnetization range is shown in red, and the minimum of the upper half
in blue. In dashed black we show the averaged magnetization (minimum
of of the free energy~\eqref{eq:FreeEnergy}) in the $K\to\infty$
limit, as a reference guide. Each point in the collection of sampled
minima was calculated by recording the average magnetization in a
histogram. This was repeated for 20 independent ISST simulations, each
of length $N=10^5$ with $\Delta t = 0.1$, whose average was used to
find the minimas. The error bars show the standard deviation of these
20 experiments.

We clearly observe in Figure \ref{fig:PhaseTrans} that the ISST
algorithm encounters no difficulties sampling the free energy surface
as the number of spins are increased. Also note that to get access to
the free energy at each temperature we simply reweigh a single ISST
trajectory \eqref{eq:reweight}, effectively creating $M=25$ copies of
the histogram, each representing the free energy at that temperature.

\subsubsection{Improvement Over Standard Simulated Tempering}
\label{sec:STvsISST}

In this section we briefly illustrate the improvement of ISST over
ST. To get an accurate comparison we implemented the ST algorithm of
Nguyen \textit{et al.} \cite{Nguyen} with adaptive weight learning. As
this method determines the weights on the fly, it only leaves us to
determine the switch frequency, switch strategy and temperature
distribution. We set the switching frequency to every timestep and
only allow for switches between consecutive temperatures either up or
down. We also distribute the temperatures linearly in $\beta$ between
$\betamin=1$ and $\betamax=3$.

\begin{figure}[!h]
	\centering
	\includegraphics{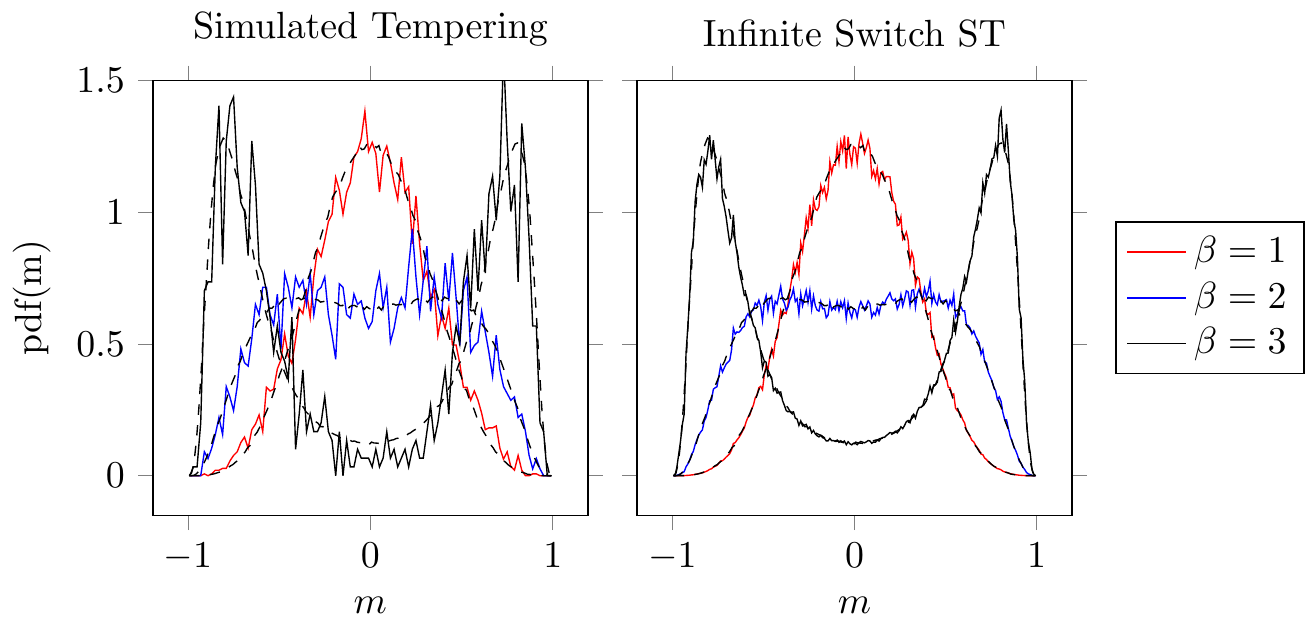}
	\caption{Sampling performance of the ISST algorithm compared
          to standard simulated tempering algorithm with adaptive
          weight learning proposing a temperature switch on every
          timestep \cite{Nguyen}. The results are shown for a Curie
          Weiss system with $K=10$ spins and external field $b=0$,
          which is simple enough to be sampled by a standard Langevin
          scheme (with results shown as reference in dashed
          black). The distribution of the average magnetization
          \eqref{eq:19} was calculated by recording a trajectory of
          length $N=10^7$ in a histogram. The ISST trajectory was
          recorded as weighted histograms from one trajectory and the
          ST trajectory was recorded in the histogram corresponding to
          the current temperature. 25 temperatures were used,
          distributed linearly in $\beta$ for ST and as the Legendre
          roots for ISST.  }
	\label{fig:STvsISST}
\end{figure}

For both methods we use 25 temperatures and we record the average
magnetization \eqref{eq:19} of a Curie Weiss system in 25 individual
histograms by recording samples from a single temperature in the case
of ST. In the case of ISST we reweigh a single trajectory. As can be
seen in Figure \ref{fig:STvsISST} this results in much better sampling
for ISST than ST. This is because ISST use the full trajectory to
compute expectations at every temperature, whereas ST only uses the
pieces of the trajectory at a given temperature to compute
expectations at that temperature. The procedure used in ISST reduces
the statistical noise significantly for the same computational cost.

\begin{figure}[!h]
	\centering
	\includegraphics{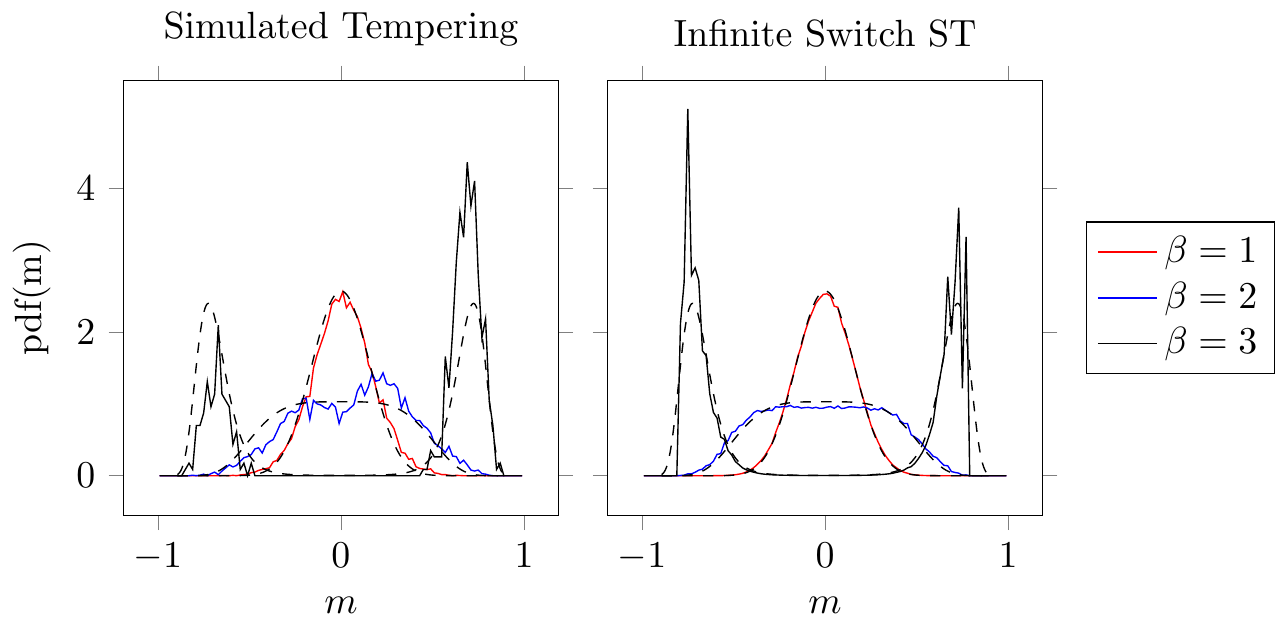}
	\caption{Difference in sampling performance between ST and
          ISST. These results are for $K=40$ Curie Weiss spins and
          external field $b=0$. The reference solution in dashed is
          the $K\to\infty$ limit result from Sec. \ref{sec:thermo}.}
	\label{fig:STvsISST2}
\end{figure}

We also performed a second experiment in which we used $K=40$ Curie
Weiss spin particles. The results of recording the average
magnetization $m$ in this case are shown in Figure
\ref{fig:STvsISST2}, where we plot the limiting result from
Sec. \ref{sec:thermo} as a guide for the eye in dashed (We therefore
do not expect perfect overlapping of the numerical experiments and the
dashed line). Again we observe the clear advantage of using the
reweighing scheme in ISST to record the statistics.

We therefore conclude that using ISST significantly improves the
sampling performance without introducing algorithmic complications for
the same computational cost. ISST also removes the need for the
practitioner to make any choices for parameters other than the limits
of the desired temperature range.

\subsubsection{Convergence of the Temperature Weights}
\begin{figure}[!ht]
	\centering
	\includegraphics{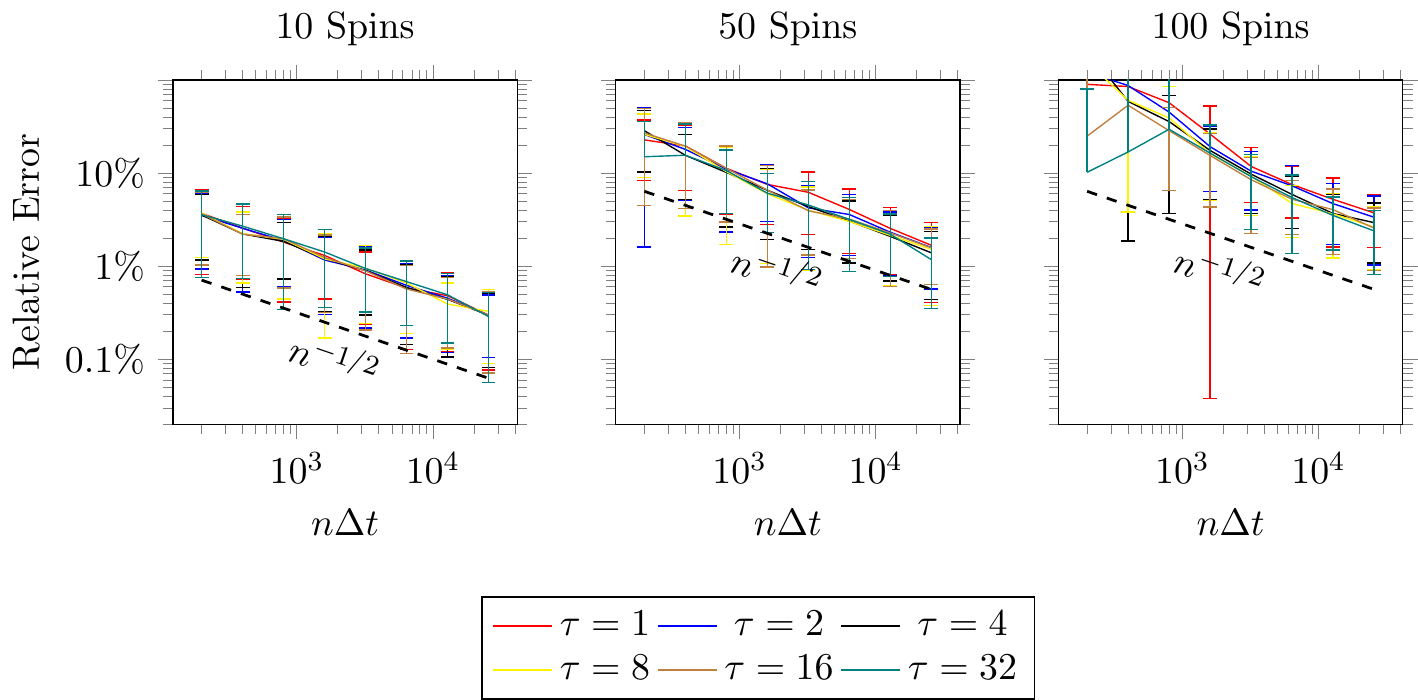}
	\caption{Convergence of the weight estimation using weight
          learning for a range of different $\tau$. The relative error
          for $n$ is calculated with respect to the weight estimate at
          $2n$ as given in \eqref{eq:RelCuEr}, i.e both the quantities
          were calculated as an average of 128 independent
          trajectories of length $n$ and $2n$ respectively. The
          relative error shown in the figures, and its standard
          deviation (error bars), was found by averaging over 128
          independent relative error estimates.}
	\label{fig:CuWeConv}
\end{figure}
Finally we recompute the experiment of Sec.~\ref{sec:CoTW}, confirming
the conclusion that $\tau$ does not play a major role in the
convergence of the temperature weights. As the long-time asymptotic
weights cannot be expressed explicitly we modify the relative error
such that,
\begin{equation}
  \label{eq:RelCuEr}
  {\rm Rel. Error} = \sup_{M > i\geq 0} \frac{\left\vert
    \mathbb{E}_{128}\left[\omega _{i,n}\right] - 
    \mathbb{E}_{128}\left[\omega_{i,2n}\right]  \right\vert}{\left\vert \mathbb{E}_{128}\left[\omega_{i,2n}\right] \right\vert}.
\end{equation}
Here, we use the notation $\mathbb{E}_{128}$ to represent an average over 128
independent ISST trajectories. We thus define the relative error as the relative difference between an average of 128
simulations of length $n$ and an average of 128 independent ISST
trajectories of length $2n$. This process is repeated 128 times to
produce the points in Figure \ref{fig:CuWeConv}, which also shows the
standard deviation of these repeated experiments as error bars.

Again, we conclude that the fixed point of the learning scheme
introduced in Sec.~\ref{sec:learning} is stable for modest values of
$\tau$. We also observe that the $n^{-1/2}$ decay of the Monte Carlo
sampling error dominates the accuracy of the adjustment scheme through
approximation of the partition functions \eqref{eq:23}.

\subsection{Alanine-12}

We implemented the ISST algorithm with adaptive weight learning in the
MIST package \cite{Bethune} and simulated the Alanine-12 molecule in
vacuum using GROMACS as host code (note: the ISST algorithm is
therefore also available to use with Amber 14, NAMD-Lite and
LAMMPS). We used 20 temperatures at the Legendre-basis in $\beta$
between 300-500K and ran the simulation for 2.2$\mu s$ with a 2f$s$
timestep. The implementation also records all the 20 individual
observable weights \eqref{eq:6}, such that the statistics can be
calculated at any desired temperature by reweighing.

\begin{figure}[!ht]
	\centering
	\includegraphics[scale=0.5]{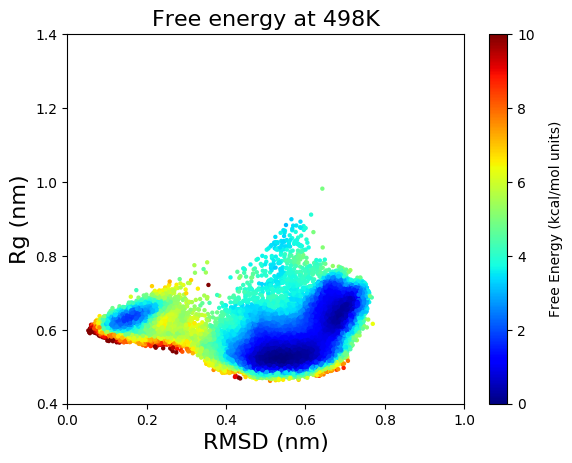}
	\includegraphics[scale=0.5]{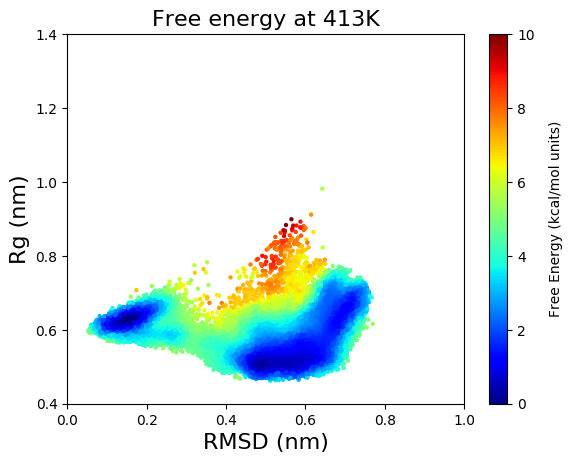}\\
	\includegraphics[scale=0.5]{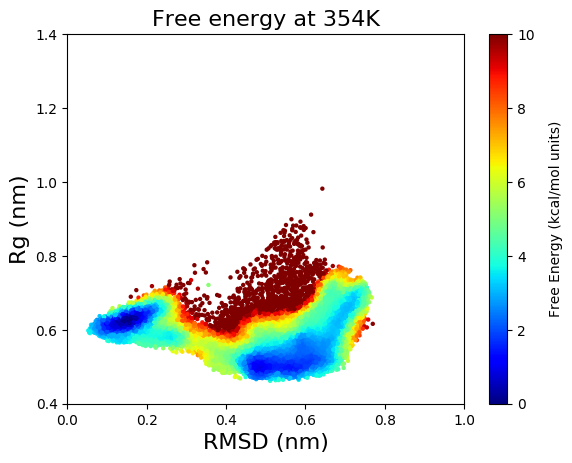}
	\includegraphics[scale=0.5]{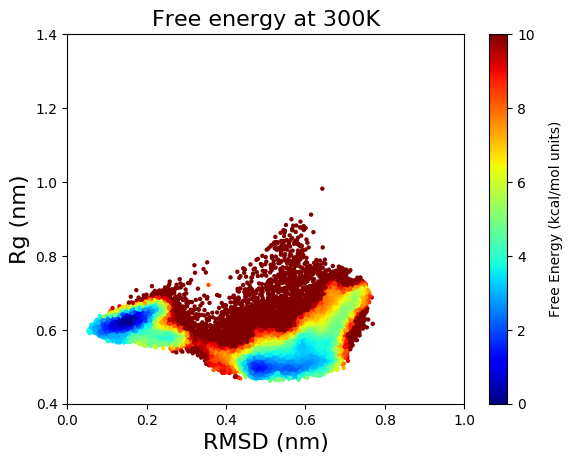}
	\caption{Free-energy obtained from an ISST trajectory for the
          in vacuum Alanine-12. The simulation started from the
          helical configuration and ran for $2.2\mu s$ using a 2 f$s$
          timestep. The Amber96 force- field was used with a
          20\AA~cut-off for electrostatics, constraining all bonds
          using the SNIP method.}
	\label{fig:alanine-12}
\end{figure}

We used GROMACS to extract the trajectory of the root mean square
deviation (RMSD) and radius of gyration $R_g$ from the initial
state. In Figure \ref{fig:alanine-12} we plot the free energy of the
RMSD and $R_g$ at four temperatures. At the high temperature (top left
panel) we observe 3 distinct states separated by energetic
barriers. As the temperature decreases towards (from top left to
bottom right) the free energy landscape changes into two distinct
states separated by a large energetic barrier. As observed in
\cite{Bethune}, initializing a vanilla MD simulation in either basin
at 300K will result in a skewed free energy landscape with no
transitions between these two states.

In contrast, by using the ISST method we obtain improved sampling at
all temperatures in the 300--500K range, both on the barriers and in
the basins. We resolve, in detail, both the shape and the
configurational states close to the states with minimal energy, giving
a good overview of the structure of the free energy landscape and the
available configurations at each temperature. Note that another
benefit of ISST is that it significantly reduces the noise in the
sampling compared to ST (this is clear from Sec.~\ref{sec:STvsISST});
ISST hence requires shorter trajectories and therefore less
computational cost to achieve satisfactory results.

\section{Concluding remarks}
\label{sec:conclu}

The theoretical analysis of the simulated tempering method that we
performed in this paper allowed us to give it a firm theoretical
justification and to draw several interesting connections between this
and other sampling techniques. First we showed, as a consequence of a
large deviation argument, that the optimal adjustment of temperature
is via infinitely frequent switches, in which case it is possible to
interpret the ST dynamics as being derived from a dynamical model for
the positions and momenta alone using a modified potential energy
function obtained by averaging.  The equations of motions used in that
model are the ones used in the integrate-over-temperature method of
Gao~\cite{Gao2008}. Second we showed that it is preferable to think of
the temperature as varying continuously in a range, in which
case ST becomes a variant of the continuous tempering method proposed
in Ref.~\cite{Gobbo2015}. Thirdly, we justified using the inverse of
the partition function as temperature weights because of its flattening effect on
the probability distribution of the system's energy, which makes ISST
the effective dual of the Wang-Landau method~\cite{WaLa2001} in this
case.

These theoretical considerations also permitted us to revisit in some
detail the implementation of the ST method within a molecular dynamics
framework. In particular we showed how to learn the partition
function, and thereby the optimal temperature weights, in a simple
adaptive way using a new estimator for this partition function.  In
our experiments on test models (harmonic oscillators, Curie-Weiss with
second order phase transition; Alanine-12 molecule in vacuum), we
found that the implementation of the ISST method was effective at
accelerating sampling in challenging situations.

\ack

We thank Iain Bethune for assistance with implementation of the ISST
algorithm within the MIST system, and Cameron Abrams, Giovanni
Ciccotti, and Glen Hocky for useful comments. J.L.~is supported in part by the
National Science Foundation (NSF) under grant DMS-1454939. B. L. is
supported in part by the Engineering and Physical Sciences Research
Council, grant EP/P006175/1.  E.V.-E.~is supported in part by the
National Science Foundation (NSF) Materials Research Science and
Engineering Center Program Award DMR-1420073; and by NSF Award
DMS-1522767. A.M. is supported by The Maxwell Institute Graduate
School in Analysis and its Applications, UK EPSRC grant EP/L016508/01.


\bigskip\bigskip


\begin{thebibliography}{10}

\bibitem{frenkel2001understanding}
Daan Frenkel and Berend Smit.
\newblock {\em Understanding molecular simulation: from algorithms to
  applications}, volume~1.
\newblock Elsevier, 2001.

\bibitem{MacKay2003InformationTI}
David J.~C. MacKay.
\newblock Information theory, inference, and learning algorithms.
\newblock {\em IEEE Transactions on Information Theory}, 50:2544--2545, 2003.

\bibitem{richard2007efficient}
Jean-Francois Richard and Wei Zhang.
\newblock Efficient high-dimensional importance sampling.
\newblock {\em Journal of Econometrics}, 141(2):1385--1411, 2007.

\bibitem{landau2014guide}
David~P Landau and Kurt Binder.
\newblock {\em A guide to Monte Carlo simulations in statistical physics}.
\newblock Cambridge university press, 2014.

\bibitem{WaLa2001}
F.~Wang and D.~P. Landau.
\newblock Efficient, multiple-range random walk algorithm to calculate the
  density of states.
\newblock {\em Phys. Rev. Lett.}, 86(10):2050–2053, 2001.

\bibitem{JuPeVo2014}
C.~Junghans, D.~Perez, and T.~Vogel.
\newblock Molecular dynamics in the multicanonical ensemble: Equivalence of
  wang–landau sampling, statistical temperature molecular dynamics, and
  metadynamics.
\newblock {\em Journal of Chemical Theory and Computation}, 10:1843--1847,
  2014.

\bibitem{LaPa2002}
A.~Laio and M.~Parrinello.
\newblock Escaping free-energy minima.
\newblock {\em Proc. Nat. Acad. Sci.}, 99(20):12562--12566, 2002.

\bibitem{BaBuPa2008}
A.~Barducci, G.~Bussi, and M.~Parrinello.
\newblock Well-tempered metadynamics: A smoothly converging and tunable
  free-energy method.
\newblock {\em Phys. Rev. Lett.}, 100(2), 2008.

\bibitem{MaVa2006}
Luca Maragliano and Eric Vanden-Eijnden.
\newblock {A temperature accelerated method for sampling free energy and
  determining reaction pathways in rare events simulations}.
\newblock {\em Chem. Phys. Lett.}, 426:168--175, 2006.

\bibitem{AbVa2010}
Cameron~F Abrams and Eric Vanden-Eijnden.
\newblock {Large-scale conformational sampling of proteins using
  temperature-accelerated molecular dynamics}.
\newblock {\em Proc. Nat. Acad. Sci. USA}, 107(11):4961--4966, 2010.

\bibitem{DaRoPo2008}
E.~Darve, D.~Rodriguez-Gomez, and A.~Pohorille.
\newblock Adaptive biasing force method for scalar and vector free energy
  calculations.
\newblock {\em J. Chem. Phys.}, 128(14), 2008.

\bibitem{Aluffi-Pentini1985}
F.~Aluffi-Pentini, V.~Parisi, and F.~Zirilli.
\newblock {Global optimization and stochastic differential equations}.
\newblock {\em Journal of Optimization Theory and Applications}, 47(1):1--16,
  1985.

\bibitem{Swendsen1986}
Rh~Swendsen and Js~Wang.
\newblock {Replica Monte Carlo simulation of spin glasses}.
\newblock {\em Phys. Rev. Lett.}, 57(21):2607--2609, 1986.

\bibitem{Geyer1991}
Charles~J Geyer.
\newblock {Markov Chain Monte Carlo Maximum Likelihood}.
\newblock {\em Computing Science and Statistics: Proceedings of the 23rd
  Symposium on the Interface}, (1):156--163, 1991.

\bibitem{Neal1996}
Radford~M. Neal.
\newblock {Sampling from multimodal distributions using tempered transitions}.
\newblock {\em Statistics and Computing}, 6(4):353--366, 1996.

\bibitem{Hansmann1997}
Ulrich~H.E. Hansmann.
\newblock {Parallel tempering algorithm for conformational studies of
  biological molecules}.
\newblock {\em Chemical Physics Letters}, 281(1-3):140--150, 1997.

\bibitem{Sugita1999}
Yuji Sugita and Yuko Okamoto.
\newblock {Replica-exchange molecular dynamics method for protein folding}.
\newblock {\em Chemical Physics Letters}, 314(1-2):141--151, 1999.

\bibitem{Marinari1992}
E~Marinari, G~Parisi, Dipartimento Fisica, Universitd Roma, Tor Vergata, and
  Sexione {Roma Tor Vergatau -Rmna}.
\newblock {Simulated Tempering: A New Monte Carlo Scheme}.
\newblock {\em Europhys. Lett. Europhys. Lett}, 19(196):451--458, 1992.

\bibitem{Hansmann1997a}
Ulrich H~E Hansmann and Yuko Okamoto.
\newblock {Numerical Comparisons of Three Recently Proposed Algorithms in the
  Protein Folding Problem}.
\newblock {\em J Comput Chem}, 18:920--933, 1997.

\bibitem{Banba2000}
Shinichi Banba, Zhuyan Guo, and Charles~L {Brooks III}.
\newblock {Efficient sampling of ligand orientations and conformations in free
  energy calculations using the lambda-dynamics method}.
\newblock {\em Journal Of Physical Chemistry B}, 104(29):6903--6910, 2000.

\bibitem{Woods2003}
Christopher~J Woods, Jonathan~W Essex, and Michael~A King.
\newblock {Enhanced Configurational Sampling in Binding Free-Energy
  Calculations}.
\newblock {\em Society}, pages 13711--13718, 2003.

\bibitem{Bitetti-Putzer2003}
Ryan Bitetti-Putzer, Wei Yang, and Martin Karplus.
\newblock {Generalized ensembles serve to improve the convergence of free
  energy simulations}.
\newblock {\em Chemical Physics Letters}, 377(5-6):633--641, 2003.

\bibitem{Okamoto2004}
Yuko Okamoto.
\newblock {Generalized-ensemble algorithms: Enhanced sampling techniques for
  Monte Carlo and molecular dynamics simulations}.
\newblock {\em Journal of Molecular Graphics and Modelling}, 22(5):425--439,
  2004.

\bibitem{FARALDO-GOMEZ2007}
Jos{\'{e}}~D. Faraldo-G{\'{o}}mez and Beno{\^{i}}t Roux.
\newblock {Characterization of Conformational Equilibria Through Hamiltonian
  and Temperature Replica-Exchange Simulations: Assessing Entropic and
  Environmental Effects}.
\newblock {\em Journal of computational chemistry}, 28:1634--1647, 2007.

\bibitem{Hritz2008}
Jozef Hritz and Chris Oostenbrink.
\newblock {Hamiltonian replica exchange molecular dynamics using soft-core
  interactions}.
\newblock {\em Journal of Chemical Physics}, 128(14), 2008.

\bibitem{Okazaki2006}
Kei-ichi Okazaki, Nobuyasu Koga, Shoji Takada, Jose~N Onuchic, and Peter~G
  Wolynes.
\newblock {Multiple-basin energy landscapes for large-amplitude conformational
  motions of proteins: Structure-based molecular dynamics simulations.}
\newblock {\em Proceedings of the National Academy of Sciences of the United
  States of America}, 103(32):11844--9, 2006.

\bibitem{Miao2014}
Yinglong Miao, William Sinko, Levi Pierce, Denis Bucher, Ross~C Walker, and
  J~Andrew Mccammon.
\newblock {Improved Reweighting of Accelerated Molecular Dynamics Simulations
  for Free Energy Calculation - Journal of Chemical Theory and Computation (ACS
  Publications)}.
\newblock {\em Journal of Chemical Theory and Computation}, 10:2677--2689,
  2014.

\bibitem{Liu2005}
Pu~Liu, Byungchan Kim, Richard~a Friesner, and B~J Berne.
\newblock {Replica exchange with solute tempering: a method for sampling
  biological systems in explicit water.}
\newblock {\em Proceedings of the National Academy of Sciences of the United
  States of America}, 102(39):13749--13754, 2005.

\bibitem{Plattner2011}
Nuria Plattner, J~D Doll, Paul Dupuis, Hui Wang, Yufei Liu, and J~E Gubernatis.
\newblock {An infinite swapping approach to the rare-event sampling problem.}
\newblock {\em The Journal of chemical physics}, 135(13):134111, oct 2011.

\bibitem{Dupuis2012}
Paul Dupuis, Yufei Liu, Nuria Plattner, and J.~D. Doll.
\newblock {On the Infinite Swapping Limit for Parallel Tempering}.
\newblock {\em Multiscale Modeling {\&} Simulation}, 10(3):986--1022, 2012.

\bibitem{Gao2008}
Yi~Qin Gao.
\newblock {An integrate-over-temperature approach for enhanced sampling}.
\newblock {\em Journal of Chemical Physics}, 128(6), 2008.

\bibitem{Gobbo2015}
Gianpaolo Gobbo and Benedict~J. Leimkuhler.
\newblock Extended hamiltonian approach to continuous tempering.
\newblock {\em Phys. Rev. E}, 91:061301, Jun 2015.

\bibitem{Park:2007id}
Sanghyun Park and Vijay~S Pande.
\newblock {Choosing weights for simulated tempering}.
\newblock {\em Phys. Rev. E}, 76(1):306, July 2007.

\bibitem{Roberts2014min}
Gareth~O Roberts and Jeffrey~S Rosenthal.
\newblock {Minimising MCMC variance via diffusion limits, with an application
  to simulated tempering}.
\newblock {\em Ann Appl Probab}, 24(1):131--149, February 2014.

\bibitem{tan2017}
Zhiqiang Tan.
\newblock Optimally adjusted mixture sampling and locally weighted histogram
  analysis.
\newblock {\em J. Comp. Graph. Stat.}, 26(1):54--65, 2017.

\bibitem{Yang:2018wy}
Yi~Isaac Yang, Haiyang Niu, and Michele Parrinello.
\newblock {Combining Metadynamics and Integrated Tempering Sampling}.
\newblock {\em arXiv.org}, July 2018.

\bibitem{Carlson:2016}
D.~E. Carlson, P.~Stinson, A.~Pakman, and L.~Paninski.
\newblock Partition functions from {R}ao-{B}lackwellized tempered sampling.
\newblock In {\em Proceedings of the $33^{\text{rd}}$ International Conference
  on Machine Learning}, volume~48. JMLR W\&CP, 2016.

\bibitem{Madras1999importance}
Neal Madras and Mauro Piccioni.
\newblock {Importance sampling for families of distributions}.
\newblock {\em Annals of Applied Probability}, 9(4):1202--1225, January 1999.

\bibitem{DonskerVaradhan:1975}
M.~D. Donsker and S.R.S. Varadhan.
\newblock Asymptotic evaluation of certain {M}arkov process expectations for
  large time, {I}.
\newblock {\em Comm. Pure Appl. Math.}, 28:1--47, 1975.

\bibitem{DeuschelStroock:1989}
J.~D. Deuschel and D.~W. Stroock.
\newblock {\em Large deviation}.
\newblock Pure and Applied Mathematics, Vol. 137. Academic Press, New York,
  1989.

\bibitem{Wu:2001}
Liming Wu.
\newblock Large and moderate deviations and exponential convergence for
  stochastic damping {H}amiltonian systems.
\newblock {\em Stoch. Proc. Appl.}, 91:205--238, 2001.

\bibitem{BodineauLefevere:2008}
T.~Bodineau and R.~Lefevere.
\newblock Large deviation of lattice {H}amiltonian dynamics coupled to
  stochastic thermostats.
\newblock {\em J. Stat. Phys.}, 133:1--27, 2008.

\bibitem{Leimkuhler}
Ben Leimkuhler and Charles Matthews.
\newblock {\em {Molecular Dynamics With Deterministic and Stochastic Numerical
  Methods}}.
\newblock Springer International Publishing, 1 edition, 2015.

\bibitem{LeMa2013b}
Benedict Leimkuhler and Charles Matthews.
\newblock {Rational Construction of Stochastic Numerical Methods for Molecular
  Sampling}.
\newblock {\em Applied Mathematics Research eXpress}, 2013(1):34--56, 2013.

\bibitem{Nguyen}
Phuong~H Nguyen, Yuko Okamoto, and Philippe Derreumaux.
\newblock {Communication: Simulated tempering with fast on-the-fly weight
  determination}.
\newblock {\em J Chem Phys}, 138(6):061102, February 2013.

\bibitem{Bethune}
Iain Bethune, Ralf Banisch, Elena Breitmoser, Antonia B~K Collis, Gordon Gibb,
  Gianpaolo Gobbo, Charles Matthews, Graeme~J Ackland, and Benedict~J
  Leimkuhler.
\newblock {MIST: A Simple and Efficient Molecular Dynamics Abstraction Library
  for Integrator Development}.
\newblock {\em arXiv}, May 2018.

\end{thebibliography}

\end{document}